\documentclass[journal=jacsat,manuscript=article]{achemso}
\usepackage{subfig}
\usepackage{chemformula}
\usepackage{graphicx}
\usepackage{dcolumn}
\usepackage{bm}
\usepackage{xcolor}
\usepackage[T1]{fontenc} 

\author{Isaac M. Craig}
\affiliation{Department of Chemistry, University of California, Berkeley, CA, USA}
\alsoaffiliation{Materials Sciences Division, Lawrence Berkeley National Laboratory, Berkeley, CA, USA}
\author{B. Junsuh Kim}
\affiliation{Department of Chemistry, University of California, Berkeley, CA, USA}
\author{David T. Limmer}
\affiliation{Department of Chemistry, University of California, Berkeley, CA, USA}
\alsoaffiliation{Materials Sciences Division, Lawrence Berkeley National Laboratory, Berkeley, CA, USA}
\alsoaffiliation{Kavli Energy NanoScience Institute, Berkeley, CA, USA}
\author{D. Kwabena Bediako}
\affiliation{Department of Chemistry, University of California, Berkeley, CA, USA}
\alsoaffiliation{Chemical Sciences Division, Lawrence Berkeley National Laboratory, Berkeley, CA, USA}
\author{Sinéad M. Griffin}
\alsoaffiliation{Materials Sciences Division, Lawrence Berkeley National Laboratory, Berkeley, CA, USA}
\email{sgriffin@lbl.gov}
\affiliation{Molecular Foundry, Lawrence Berkeley National Laboratory, Berkeley, CA, USA}
\title[title]{ Modeling the Superlattice Phase Diagram of Transition Metal Intercalation in Bilayer \textit{2H}-TaS$_2$ }

\begin{document}
\newcommand{\three}{{$\sqrt{3} \times \sqrt{3}$} }
\newcommand{\two}{{$2 \times 2$} }

\begin{abstract}
Van-der-Waals hosts intercalated with transition-metal (TM) ions exhibit a range of magnetic properties strongly influenced by the structural order of the intercalants. However, predictive computational models for the intercalant ordering phase diagram are lacking, complicating experimental pursuits to target key structural phases. Here we use Density Functional Theory (DFT) to construct a pairwise lattice model and Monte Carlo to determine its associated thermodynamic phase diagram. To circumvent the complexities of modeling magnetic effects, we use the diamagnetic ions Zn$^{2+}$ and Sc$^{3+}$ as computationally accessible proxies for divalent and trivalent species of interest (Fe$^{2+}$ and Cr$^{3+}$), which provide insights into the high-temperature thermodynamic phase diagram well above the paramagnetic transition temperature. We find that electrostatic coupling between intercalants is almost entirely screened, so the pairwise lattice model represents a coarse-grained charge density reorganization about the intercalated sites. The resulting phase diagram reveals that the entropically-favored \three ordering and coexisting locally ordered \three and \two domains persist across a range of temperatures and intercalation densities. This occurs even at quarter filling of interstitial sites (corresponding to bulk stoichiometries of \textit{M}$_{0.25}$TaS$_2$; \textit{M} = intercalant ion) where long-range \two order is typically assumed. 
\end{abstract}

\section{Introduction}
Van-der-Waals (vdW) materials intercalated with transition metal (TM) ions offer a large phase space for novel magnetic systems, realizing a broad array of technologically advantageous properties.~\cite{hulliger_magnetic_1970,van_laar_magnetic_1971, DiSalvo1973, Parkin1980,friend1977electrical, rao1979, Wilson1969, marseglia1983, Parkin1980a,EIBSCHTZ1981, Maksimovic_et_al:2022,Togawa2012} The ability to independently adjust the host lattice, intercalant chemical identity, stoichiometry, and intercalant ordering within the host enables huge control over the resulting spin Hamiltonian and emergent functionality. These include large magnetic anisotropies\cite{Morosan2007} that stabilize magnetic order down to the 2D-limit \cite{husremović2022hard}, large Dzyaloshinskii–Moriya interactions promoting chiral magnetic textures \cite{Miyadai1983, Togawa2012, zhang2021chiral, ghimire_magnetic_2013}, switches between distinct resistive states with remarkably low current densities, \cite{nair2020electrical, maniv2021antiferromagnetic} among others. For all of these cases, the sites the intercalants reside in, their densities, and the resultant longer-range "superlattices" that form are key to understanding their diverse magnetic properties \cite{wu2022highly, li2023role, dyadkin2015structural, weber2021origins, goodge2023consequences, xie2022structure}.

Among these, intercalated \textit{2H}-TaS$_2$ shows a range of magnetic behavior strongly dependent on its superlattice order. When 1/4 and 1/3 of the octahedral interstitial sites are occupied in a nearly ordered manner (corresponding to bulk compositions of \textit{2H}-$M_{0.25}$TaS$_2$ and \textit{2H}-$M_{0.33}$TaS$_2$ respectively, where $M$ = Fe, Cr), the intercalants can form \two and \three superlattices, with distinct magnetic properties. ~\cite{hulliger_magnetic_1970, van_laar_magnetic_1971, Parkin1980a} The magnetic behaviors of these stoichiometric compounds differ significantly from the disordered phases typically found in off-stoichiometric samples \cite{Hardy2015, Horibe2014, Chen2016f, maniv2021exchange, zhang2019critical, kong2023near}. This observation is often attributed to oscillatory exchange mechanisms~\cite{ko2011rkky, rahman2022rkky} such as Ruderman–Kittel–Kasuya–Yosida (RKKY) exchange,~\cite{kasuya1956theory, yosida1957magnetic, ruderman1954indirect}, in which the magnitude and sign of magnetic exchange constant depend on the intercalant spacing. Because of this, subtle changes in intercalant ordering can have a pronounced effect on the observed magnetism. For instance, \textit{2H}-Fe$_x$TaS$_2$ displays ferromagnetic or antiferromagnetic behavior depending on the intercalation density, $x$.~\cite{Chen2016f, Morosan2007, EIBSCHTZ1981, NARITA1994, Mangelsen2020} Similarly, subtle variations in magnetic order have been used to explain the switching behavior in Fe$_x$NbS$_2$, where the magnetic properties are significantly influenced by the precise intercalation density of the sample. \cite{wu2022highly, weber2021origins, maniv2021antiferromagnetic, nair2020electrical, maniv2021exchange} Notably, angle-resolved photoemission spectroscopy suggests that the magnetic exchange in these systems may not always be adequately described by the standard RKKY mechanism \cite{sirica2020nature, xie2023comparative}, highlighting the need for further studies into the interplay between superlattice order and magnetism in these systems. Despite extensive modeling of superlattice order in alkali metal intercalation compounds using DFT-informed lattice Hamiltonians, \cite{lee2012li, persson2010thermodynamic, van2008nondilute, emly2015mg, wolverton1998first, van1998first, koudriachova2002density} computational predictions for superlattice order in TM intercalated TMDs remain absent. A key implication of this knowledge gap has led to the conventional assumption that specific compositions (especially stoichiometric) strongly favor certain superlattice structures -- for example, $x=1/4$ is thought to imply \two ordering, and $x=1/3$ implies \three. In this work, we show this assumption is incorrect for $x=1/4$. 

While the intercalant ordering strongly influences the observed magnetic properties, the magnetic phenomena typically occur at much lower temperatures ($\sim$10-150 K) than the solid-state synthesis and crystal growth processes ($\sim$1000-1400 K). \cite{hulliger_magnetic_1970, van_laar_magnetic_1971, Parkin1980a, husremović2022hard} This suggests that the magnetic coupling does not play a significant role in determining the site occupancy energies at temperatures well above the paramagnetic transition. While this supports a decoupled treatment of magnetic and structural effects when interpreting experimental results, it complicates the design of informative calculations as conventional ground-state electronic structure calculations, without specialized treatment, require incorporating all of the many competing magnetic and structural effects. \cite{abrikosov2016recent} To circumvent these challenges, we focus on two intercalants -- Zn and Sc -- that serve as proxies for the magnetic ions of interest (Fe and Cr) but are nominally diamagnetic. Zn and Sc are expected to form 2+ and 3+ oxidation states, respectively, following charge transfer with the host lattice, similar to Fe and Cr. However, the work function of Sc (3.5 eV) is slightly lower than that of Cr (4.5 eV), suggesting increased charge transfer for Sc. \cite{michaelson1977work} Similarly, the work function of Zn (4.33 eV) is also slightly smaller than that of Fe (4.5 eV). \cite{michaelson1977work} Both Zn and Sc have ionic radii (0.74 \textrm{\AA} and 0.75 \textrm{\AA}, respectively) comparable to those of Fe and Cr (0.78 \textrm{\AA} each).\cite{shannon1976revised} We therefore expect that Fe and Cr will have similar structural and charge transfer behavior to Zn and Sc. The chosen host material, \textit{2H}-TaS$_2$, supports predominantly octahedral interstitial site occupation \cite{van_laar_magnetic_1971}, allowing for a relatively simple description of possible intercalation sublattices. 

Here, we investigate the equilibrium distributions of Zn and Sc intercalations in bilayer \textit{2H}-TaS$_2$. Using Monte Carlo simulations based on a lattice Hamiltonian derived from Density Functional Theory (DFT), we map the lattice occupancy phase diagram of these representative transition metal ions within this commonly chosen vdW host. Our findings show that the large dielectric of this metallic host screens the electrostatic interactions between interstitial ions almost entirely, even for nearest-neighbor interactions. As a result, the lattice Hamiltonian primarily captures the reorganization of bonding orbital density, rather than long-range electrostatic coupling. This short-range interaction, coupled with relatively large inter-site distances, leads to rapidly decaying interactions that preclude a mean-field treatment and largely disfavor ion clustering. Consequently, the preferences for \two and \three ordering are driven predominantly by configurational entropy. The resulting phase diagram reveals that the \three order persists over a much broader range of intercalation densities and temperatures than the $2 \times 2$, even in compositional regimes where \two ordering might be conventionally expected (such as the stoichiometric bulk compositions of $x=0.25$). Instead, the \two superlattice order is confined to a narrow region of the phase diagram. By considering the intercalant order within a rigid bilayer host, our model represents confined crystallization within a fixed 2D lattice, where long-range order can persist is mathematically equivalent \cite{lee1952statistical} to long-range order between discrete Ising spins. \cite{mermin1966absence, hohenberg1967existence, halperin2019hohenberg}. These results help establish a framework for modeling more complex intercalated TMD systems and inform experimental pursuits to tune superlattice order with thermal post-processing.  

\begin{figure*}
\includegraphics[width=16cm]{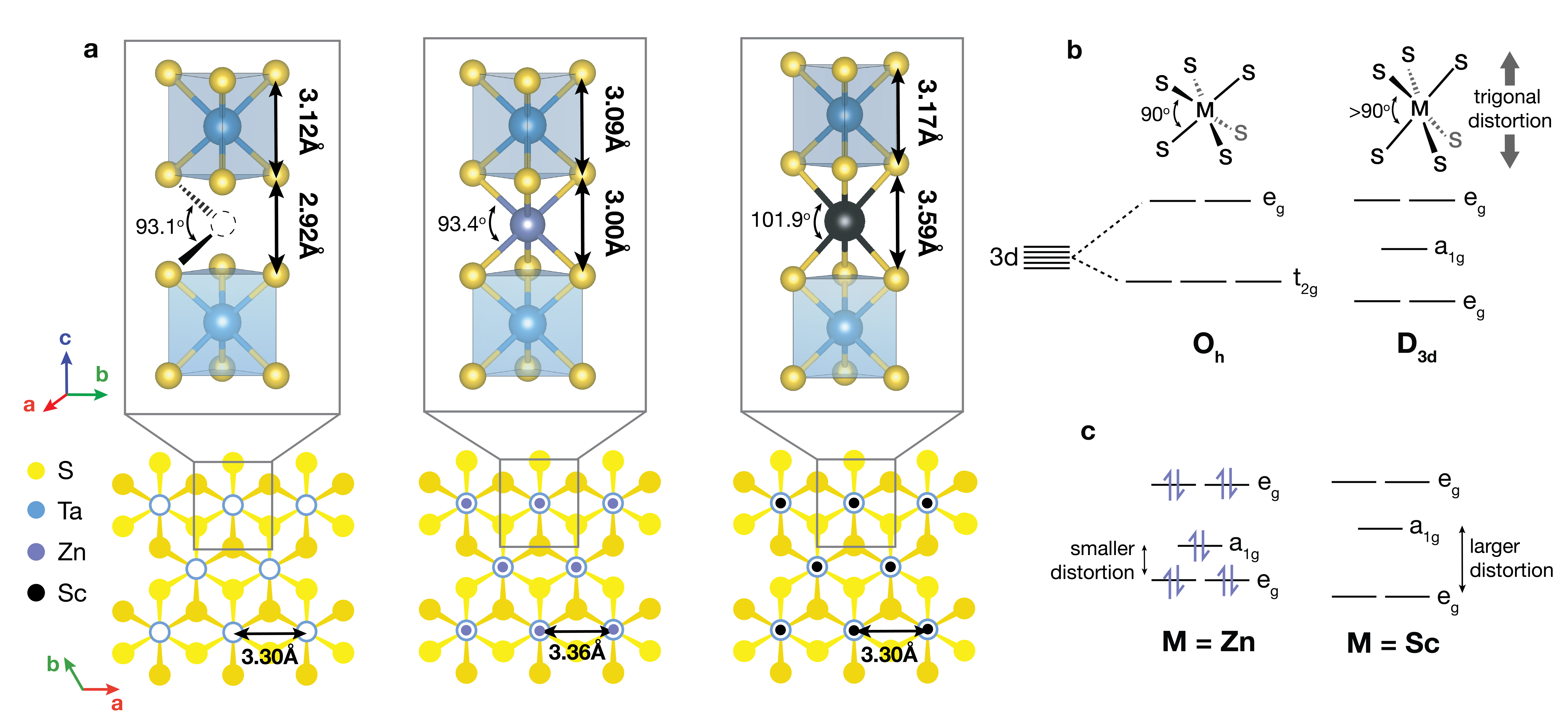}
\caption{ (a) DFT-obtained structural parameters obtained for bilayers of \textit{2H}-TaS$_2$, \textit{2H}-Zn$_{0.5}$TaS$_2$ and \textit{2H}-Sc$_{0.5}$TaS$_2$. \textit{2H}-M$_{0.5}$TaS$_2$ structures correspond to the occupation of all pseudo-octahedral interstitial sites within the vdW gap of this bilayer. A trigonal distortion, manifesting as a bond angle greater than $90^{o}$, is present even in the vacant TaS$_2$. \textbf{(b)} Impact of the trigonal distortion on the crystal field splitting. \textbf{(c)} Anticipated \textit{d} orbital electron configurations for Zn$^{2+}$ and Sc$^{3+}$.  } 
\end{figure*}

\subsection{Computational Methods}

We begin by constructing an extended pairwise lattice model to describe the occupational Hamiltonian. Lattice models have been used extensively to model intercalation compounds, ranging from simple pairwise models \cite{persson2010thermodynamic, kirczenow1985domain, cai1987lattice} to more complex higher-order cluster models \cite{de1994cluster, sanchez1984generalized, van2002effect, puchala2013thermodynamics, thomas2013finite, hadke2019effect, natarajan2017symmetry, lee2012li, van1998first, wolverton1998first, cordell2021probing} which account for interactions between three or more ions. We define the occupation variables $\sigma_i \in \{0, 1\}$ for each lattice site $i$, where $\sigma_i = 0$ and $\sigma_i = 1$ represent a vacant and intercalated interstitial octahedral site respectively, forming a triangular lattice. The energy of an intercalated lattice $E(\sigma)$ is given below, where $E_0$ is the vacant lattice energy, $J(r)$ is the distance-dependent pairwise coupling between intercalants separated by a distance $r_i$, and $\mu$  is the reference intercalant chemical potential (relative to the vacant lattice). We demonstrate that the accuracy of this pairwise approach varies
depending on the specific intercalant. For higher oxidation state ions, cluster models become increasingly important to capture the full complexity of the interactions. The resultant Hamiltonian corresponds to a 2D triangular lattice gas model with interactions extending beyond nearest-neighbor coupling, \cite{landau1983critical, bartelt1984triangular} 
 \begin{align*}
E(\sigma) = E_0 + \sum_{ij} J(|r_i-r_j|) \sigma_i \sigma_j + \mu \sum_i \sigma_i
\end{align*}

where the first sum is over all pairs of sites (we follow this convention for easier comparison with prior elasticity work\cite{frechette2019consequences} and fitting, rather than summing, over all unique site pairs $i > j$). The values for $J(|r_i-r_j|)$ are obtained by fitting to DFT results based on a series of ion occupancies, with further fitting details in the SI. 
  DFT calculations are performed at the GGA+U level using a Dudarev-type U \cite{dudarev1998electron} obtained from linear response \cite{kulik2006density, cococcioni2005linear}. All DFT calculations are carried out on a rigid lattice throughout, justified by the relatively small elastic energy contribution, which will be discussed further in a later section. We use the Kawasaki Algorithm to perform equilibrium Monte Carlo calculations on the occupation variables, in which randomly selected lattice sites are exchanged. Kawasaki dynamics are a common choice for performing simulations at fixed intercalation densities. Additional computational details are available in the SI. 

We differentiate superlattice phases with the following order parameters. The sublattice intercalation density, $f_{\omega}$, is defined as the average occupancy for all of the $N_\omega$ sites $i$ within a sublattice $\omega$, $f_{\omega} = \frac{1}{N_\omega} \sum_{i \in \omega} \sigma_i$. \cite{landau1983critical} To identify \two ordering, we divide the lattice into four distinct  \two sublattices. For a rigid lattice, the location of each lattice site $i$ can be expressed using the lattice vectors $a_1$ and $a_2$ and two integer coefficients, $n_i a_1 + m_i a_2$. The four distinct \two sublattices can therefore be easily identified as having $n_i$ and $m_i$ both even, $n_i$ and $m_i$ both odd, $n_i$ even and $m_i$ odd, or $m_i$ even and $n_i$ odd provided that the total lattice dimension is divisible by 2. A similar process can be applied to identify the three distinct \three sublattices. To quantify long-range order, we define multiple order parameters $\gamma_i$ for the \three sublattices and $\phi_i$ for the \two sublattices.  For instance, the order within the three different \three sublattices are distinguished using $\gamma_1 = \frac{2f_1 - f_2 - f_3}{4}$, $\gamma_2 = \frac{2f_2 - f_1 - f_3}{4}$, and $\gamma_3 = \frac{2f_3 - f_1 - f_2}{4}$. \cite{landau1983critical} Similarly, four order parameters $\phi_i$ are defined for the \two sublattices, including $\phi_1 = \frac{3f_1 - f_2 - f_3 - f_4}{6}$ and its three other permutations. The root mean square (RMS) of the $\gamma_i$ and $\varphi_i$ is then a measure of this order. Random or uniformly distributed intercalants within each sublattice result in RMS values of zero. Perfect \three ordering corresponds to an $RMS(\gamma_i)$ of $\sqrt{3/8}$ and an $RMS(\varphi_i)$ of zero and perfect \two ordering corresponds to an $RMS(\gamma_i)$ of zero and an $RMS(\varphi_i)$ of $\sqrt{1/3}$. 

We also define a local order parameter, $\alpha_i$, following previous work \cite{cowley1950approximate, mirebeau1984first, erhart2008short}, to compare with experimental data with finite probe width and to analyze the locally ordered phases that emerge due to strong disfavoring of nearest-neighbor occupation. This short-range order parameter $\alpha_i$ for a given intercalated site $i$ is defined by $\alpha_i = 1 - \frac{N_{vac}^{(i)}}{6(1-f)}$ where $N_{vac}^{(i)}$ is the number of vacancies in the six nearest neighboring sites and $f$ is the overall fraction of occupied interstitial sites. $\alpha_i$ is one when all of these nearest neighbors are occupied, zero for a random distribution of intercalants, and reaches its minimum value of $-f/(1-f)$ when all nearest neighbors are empty. $\alpha_i$ is computed for each intercalated site and averaged to obtain the presented $\bar{\alpha}$. For bilayer interfaces, the fraction of occupied sites $f$ is twice the stoichiometric ratio $x$ in M$_{x}$TaS$_2$, whereas for bulk (3D) crystals, $x=f$.

\begin{figure*}
\centering
{\includegraphics[width=7.7cm]{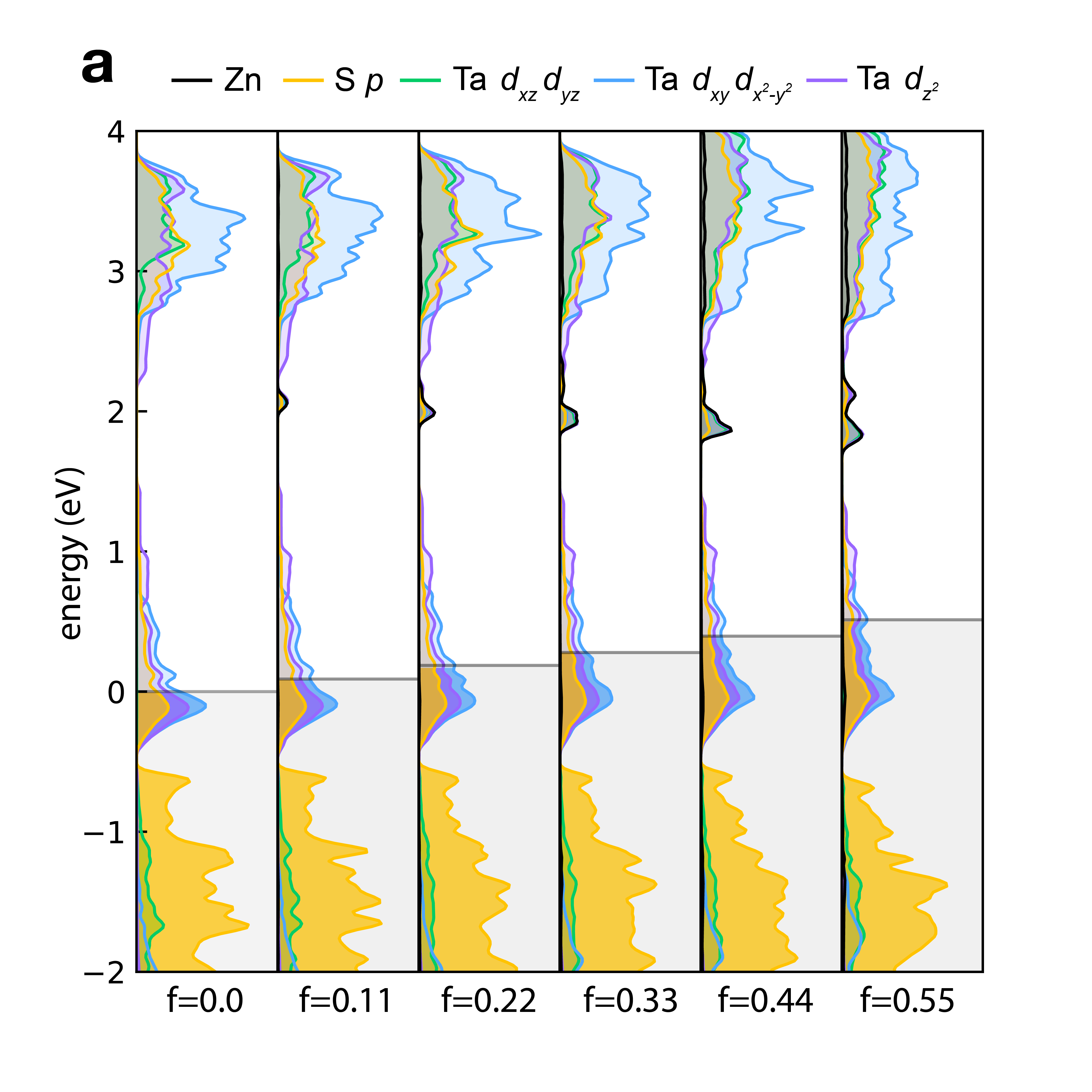} }
\qquad
{\includegraphics[width=7.7cm]{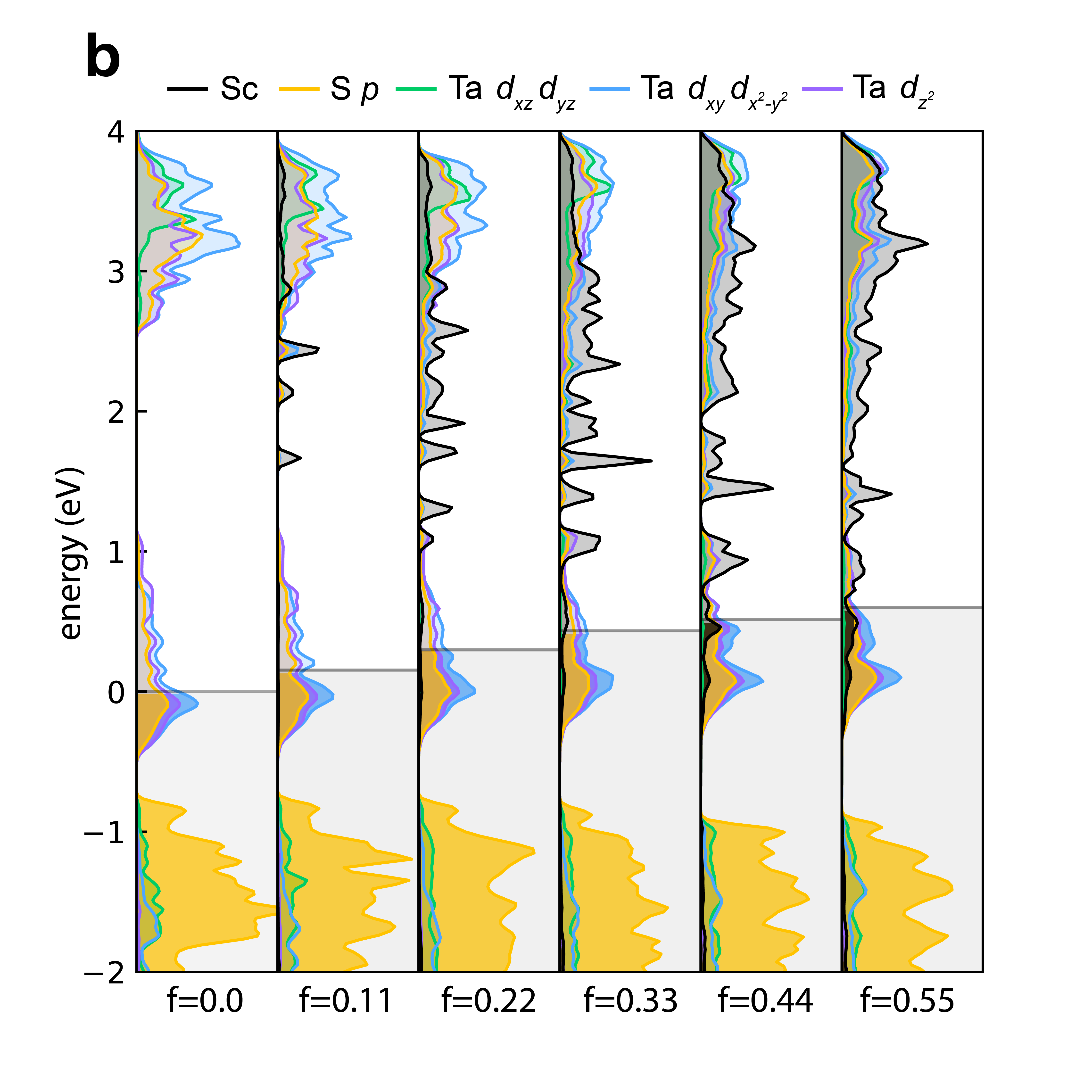} }
\caption{ Calculated electronic density of states (DOS) for bilayers of (a) \textit{2H}-Zn$_{x}$TaS$_{2}$ and (b) \textit{2H}-Sc$_{x}$TaS$_{2}$ at a series of intercalation densities, $x$. The vertical axis values are the total energies of a $3 \times 3$ bilayer units (\textit{2H}-M$_{9x}$Ta$_{18}$S$_{36}$), and are plotted relative to the Fermi energy at $x=0$ set to zero. Fermi energies at $x\neq 0$ are shown as grey lines. The x and z axes are defined parallel to a and c respectively, as shown in Fig. 1. The two nearly degenerate orbital sets $(d_{xz}, d_{yz})$ and $(d_{x^2-y^2}, d_{xy})$ are summed. Plots showing projection onto spherical harmonics using the radii from a Bader charge analysis show similar qualitative results to this projection onto the PAW sphere (see SI). The DOS at different occupation orders are shown in the SI and illustrate a comparable degree of hybridization throughout. } 
\end{figure*}

\section{Results and Discussion}
\subsection{Intercalation-driven Structural Effects}
We first investigate the anticipated extent of structural relaxation of the TaS$_2$ host lattice around the intercalated ions at the GGA+U level. Assessing this structural relaxation is important before adopting a rigid lattice model as previous work demonstrated that elastic effects from phase mismatch in multiple-component systems -- mathematically analogous to expansion about intercalants in this system -- can drive the formation of a \two ordered phase. \cite{frechette2019consequences} We aim to justify the omission of in-plane elastic effects, though not necessarily out-of-plane effects, for Zn and Sc-intercalated \textit{2H}-TaS$_2$, and to evaluate the impact of the observed structure on the magnetic and electronic landscape. 

Following structural relaxation of a maximally intercalated bilayer slab ($x=0.5$, $f=1.0$), we find that full Zn intercalation drives a small increase in both the in-plane and out-of-plane lattice constants $a_0$ (2\% change) and $c$ (3\% change), resulting in a S--Zn length of 2.45 Å. For Sc-intercalated TaS$_2$ ($x=0.5$, $f=1.0$), we see minimal change in $a_0$ and a much larger expansion out-of-plane (23\% change) for a S--Sc distance of 2.62 Å. Further structural details are available in the SI. The behavior of Zn-intercalated \textit{2H}-TaS$_2$ is consistent with previous findings of the 0.2\% and 4.6\% expansions in $a_0$ and $c$ observed upon intercalation in bulk \textit{2H}-Zn$_{0.5}$TaS$_2$, \cite{di1973metal} and also aligns with the marginal $a_0$ and $c$ expansions seen with increasing $x$ in \textit{2H}-Fe$_x$TaS$_2$. \cite{van_laar_magnetic_1971, zhang2019critical} Similarly, smaller in-plane lattice expansions have been reported in \textit{2H}-Cr$_x$TaS$_2$, compared to  its Fe-intercalated counterpart \cite{parkin19803, friend1977electrical}, consistent with the smaller $a_0$ expansion predicted here for Sc- than Zn-intercalated \textit{2H}-TaS$_2$. While there are no reports of Sc-intercalated \textit{2H}-TaS$_2$ that we are aware of, the large c-axis increase in \textit{2H}-Sc$_x$TaS$_2$ is somewhat surprising, given that 3+ intercalants are often associated with smaller c-axes than their 2+ counterparts. \cite{parkin19803, friend1977electrical}. Nevertheless, we emphasize that these comparisons are made at different occupational fractions (as $f=1$ has not been experimentally accessible) and, as we will show, these structural considerations play a relatively minor role in the intercalant ordering. 

The relatively minor changes to $a_0$ for both Zn and Sc intercalation suggest that in-plane elastic effects can be neglected in the modeling of these 2D lattices. The effective elastic Hamiltonian describing these in-plane elastic effects is proportional to $K\Delta^2/8$ where $K$ is the spring constant and $\Delta$ is proportional to the change in $a_0$ between an intercalated and vacant lattice (see SI for details and values). \cite{frechette2019consequences} The elastic effects therefore occur on an energy scale of $\sim 0.4$ meV in this system, which is several orders of magnitude smaller than the dominant electronic effects (with $J(a_0) \sim 200-300$ meV as will be shown in a later section).  For the elastic energy contribution to approach the penalties associated with next-nearest neighbor interactions, $J(\sqrt{3}a_0) \sim 20$ meV, the intercalant-induced expansion would need to be roughly seven times larger—well beyond what experimental values suggest. While elastic interactions decay more slowly than the electronic effects, their cumulative impact remains relatively minor. Monte Carlo simulations using this elastic Hamiltonian predict ordering below $T \approx 1$ in units of $\epsilon/k_b$ where $k_b$ is the Boltzmann constant. \cite{frechette2019consequences} The obtained energy scale of $\epsilon \approx 0.4$ meV (see SI) implies elasticity-driven ordering below $T \approx 4.6 K$, which has a negligible effect on the phase diagram presented here, as the ordering transitions occur at temperatures 2 to 3 orders of magnitude higher. 

\subsection{Intercalation-driven Electronic Effects }

We next examine the changes to the electronic density of states (DOS) following intercalation. This has been commonly understood \cite{di1973metal, xie2022structure} and modeled \cite{hatanaka2023magnetic} using a rigid band picture, in which charge transfer occurs from the intercalant to the host lattice \textit{d} states with minimal hybridization. This approach has important implications for the validity of the assumed RKKY-type coupling between ions. \cite{kasuya1956theory, yosida1957magnetic, ruderman1954indirect}. 

We present DOS at representative intercalation densities for Zn (Fig. 2a) and Sc (Fig. 2b) intercalated \textit{2H}-TaS$_2$ bilayers.  The DOS shows only marginal differences between different ion orderings at the same intercalation density (see SI), with the most notable variation occurring in the sulfur bands. As seen in Fig. 2a, the rigid band picture holds reasonably well at low intercalation densities, as the Zn states remain largely isolated from the host states, with minimal hybridization. This contrasts with what has been observed in DFT studies of Fe$_x$TaS$_2$ \cite{fan2017electronic, ko2011rkky, mankovsky2015electronic}, where more hybridization is evident, likely due to the smaller size of Zn orbitals.

The orbital character of the states is consistent with crystal field analysis, corroborating experimental results\cite{Wilson1969, Bell1976}. We find that the states around the Fermi level consist primarily of Ta $d_{z^2}$ character and the highest shown states primarily of $(d_{x^2-y^2}, d_{xy})$ and $(d_{xz}, d_{yz})$. While the DOS reflects relatively low hybridization overall, subtle changes are observed in the higher-energy states with predominant Ta $d$ character, which span a broader energy range as the intercalation density increases, particularly when more than one-third of the octahedral interstitial sites are occupied ($x > 0.167$). We note that this projection onto PAW spherical harmonics provides only a qualitative picture of the orbital character, comparable to results obtained using spherical harmonics with radii from a Bader charge analysis (see SI Fig. 1). A more detailed Wannier orbital analysis is not required for this qualitative picture.

The steady increase in the Fermi level reflects the expected charge transfer from the Zn states to the  Ta $d_{z^2}$ states. The filling of this band, from half-filled to nearly full, suggests a doping level of approximately one electron per Zn ion, consistent with the calculated Bader charges of 0.90 and 0.83 for $x=1/18$ and $x=0.5$, respectively (corresponding to the occupation of one-ninth and all octahedral interstitial sites). While DFT-calculated Bader charges typically underestimate oxidation values, \cite{posysaev2019oxidation} the extent of this underestimation is smaller than what we observed here, indicating that the Zn ions exhibit more 1+ oxidation state character. Additionally, the obtained Bader charges are only weakly affected by the choice of U, as U value choices of 0, 1, 2, and 3 yield Bader charges of 0.836, 0.832, 0.828, and 0.825 respectively at full occupancy $f=1$. 

For Sc-intercalated TaS$_2$, the rigid band picture breaks down more significantly, similar to the large Cr-Ta hybridization seen in Cr-intercalated TaS$_2$ \cite{xie2023comparative}. This more dramatic deviation from the rigid band assumption for Sc-intercalated TaS$_2$ can be attributed to the larger size of the Sc orbitals and the closer energy alignment between Sc and S, leading to increased hybridization. The reduced charge transfer expected from this hybridization is reflected in the calculated Bader charges of 1.86 and 1.47 for $x=1/18$ ($f=1/9$) and $x=0.5$ ($f=1$), respectively. The calculated Bader charges show some dependence on the choice of U, with U values of 0, 1, 2, and 2.9 yielding Bader charges of 1.341, 1.386, 1.431, and 1.471 respectively at $f=1$. Although this U dependence is larger for Sc than for Zn, it remains relatively small overall. The decreased charges observed for both Zn and Sc arise from the inclusion of diffuse host charge density which remains spatially localized around the intercalant, as seen in the following section. 

\begin{figure}
\includegraphics[width=8cm]{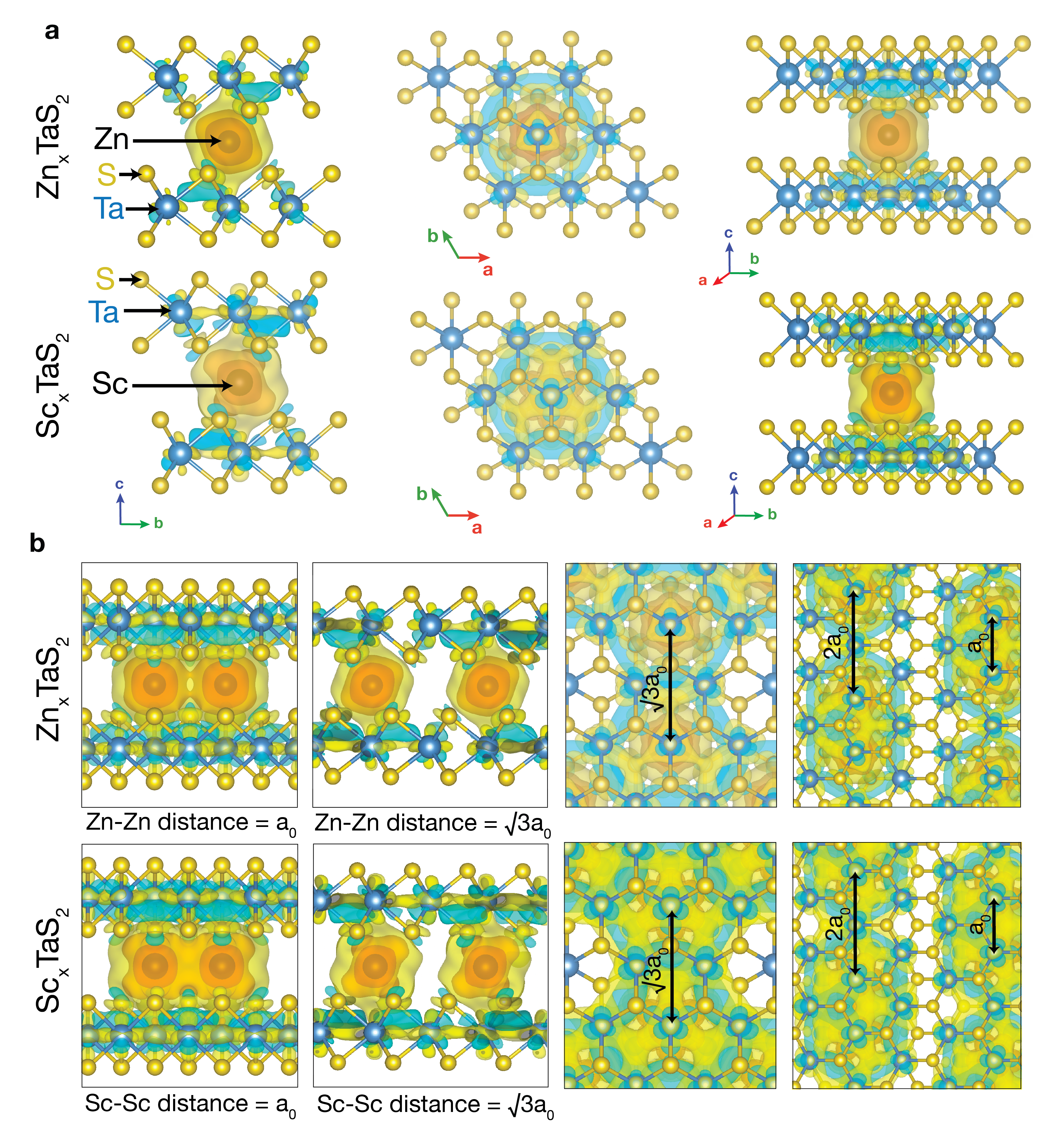}
\caption{ Charge density difference of bilayer \textit{2H}-TaS$_2$ following dilute ($x=1/18$) Sc intercalation (a) and Zn intercalation (b), in which one-ninth of the octahedral interstitial sites are occupied such that no interactions at distances $r < 3a_0$ are present. Red, orange, yellow, and cyan isosurfaces reflect values of 0.025, 0.01, 0.0015, and -0.0015 a.u. per Bohr radius cubed respectively. (b) Charge density differences for representative non-dilute occupations. 
} 
\end{figure}

\subsection{Lattice Hamiltonian Fit and Interpretation}

We next examine the spatial extent of the electron density changes following intercalation of Zn and Sc. After charge transfer, the electrons that dope the host remain predominantly localized in a diffuse shell around the Zn, rather than dispersing throughout the lattice (Fig. 4). This local modification to the charge density around the Zn ions manifests as a large increase in charge in the vdW interface and relatively minor modifications to the TMD lattice charge distribution (on the order of 0.0015 a.u. per Bohr radius cubed) within a radius of roughly $a_0$ from the intercalated site (Fig. 3a). 

Qualitatively similar behavior is seen for Sc-intercalated TaS$_2$ (Fig. 3a), with the electron density changes marginally more dispersed, as expected due to the larger extent of charge transfer. Fig. 3b indicates that interactions $J(r \geq 2a_0)$ are associated with negligible charge density modifications and therefore will not play a large role. Likewise, the charge density plots suggest that $J(r = \sqrt{3}a_0)$ interactions are driven primarily by the charge modifications within the surrounding TaS$_2$ lattice. The charge density differences following intercalation (Fig. 3a) are indistinguishable from those obtained using smaller U values, which only show differences on the order of 0.0001 a.u. per cubic Bohr radius (see SI).

\begin{figure}
\includegraphics[width=8cm]{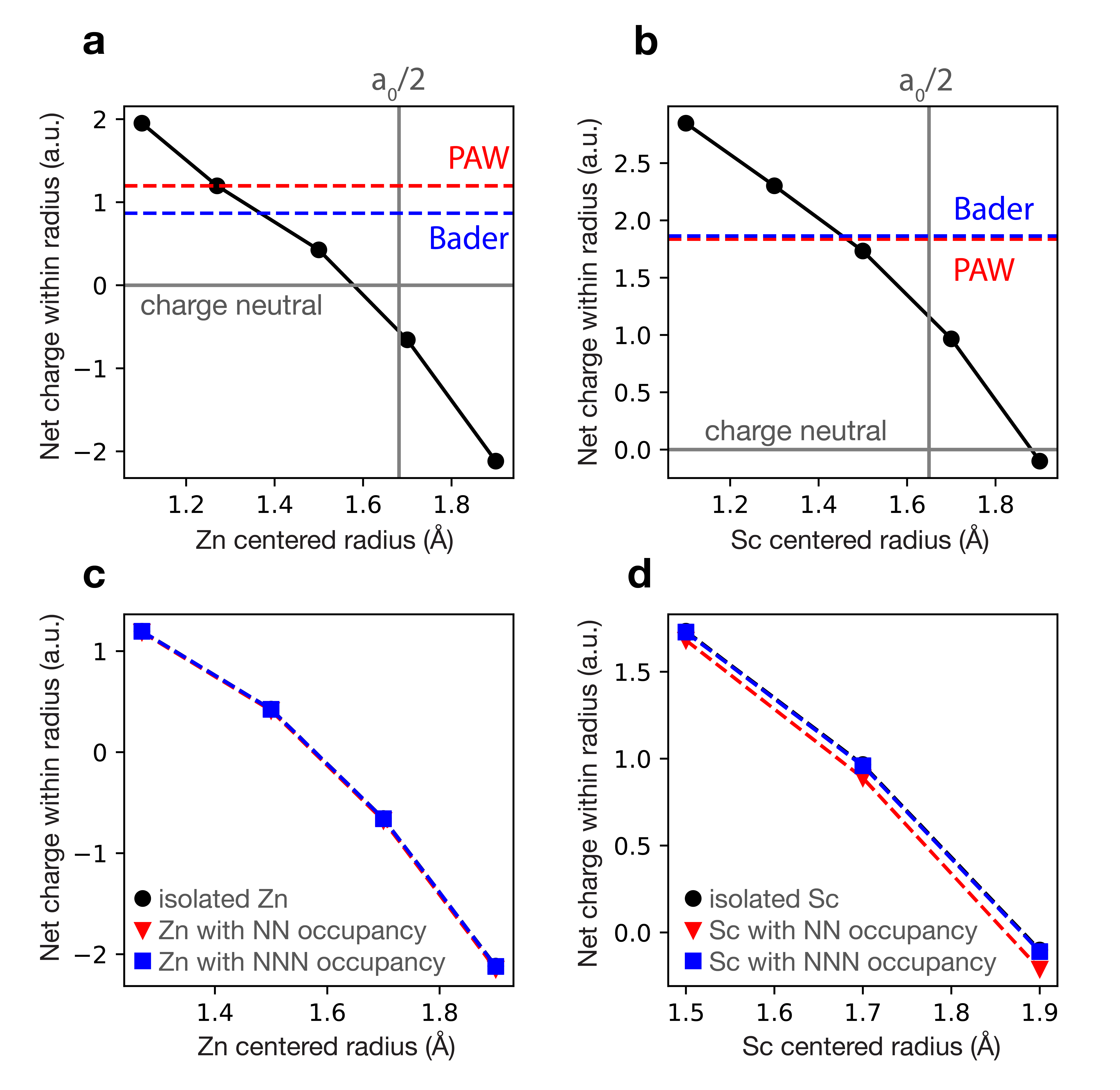}
\caption{ Total charge within spheres of increasing radii centered on the Zn (a) and Sc (b) intercalant sites. The total charge within the PAW sphere and the Bader charge are labeled. Results are shown for dilute intercalation $x=1/18$ such that no interactions at distances $r < 3a_0$ are present (a-b). Comparison of the total charge within spheres of increasing radii centered Zn (c) and Sc (d) for a series of intercalant configurations, including the dilute configuration (black) shown in subplots (a-b), an $x=2/18$ configuration in which each intercalant has three next nearest neighbor (NNN) $J(\sqrt{3}a_0)$ interactions (blue), and an $x=2/18$ configuration in which each intercalant has one nearest neighbor (NN) $J(a_0)$ interaction (red).
} 
\end{figure}
To quantify the screening from this local charge transfer, we compute the total charge contained within spheres of increasing radii centered on the Zn and Sc sites. We find that the Zn and Sc ions are entirely charge compensated at radii of 1.58 \textrm{\AA} and 1.88 \textrm{\AA} respectively (Fig. 4a-b). The presence of nearby intercalated sites has negligible impact on the observed trend for Zn-intercalated TaS$_2$ (Fig. 4c) and only a minor impact for Sc-intercalated TaS$_2$ (Fig. 4d). The fact that the Zn and Sc ions are charge compensated by 1.58 \textrm{\AA} and 1.88 \textrm{\AA} respectively implies that electrostatic coupling between Zn ions and between Sc ions are completely screened for $J(r \geq a_0)$ and $J(r \geq \sqrt{3} a_0)$ respectively.  These charge modifications imply that the coupling constants $J(r)$ are not primarily driven by electrostatic interactions but instead represent a coarse-grained picture of the local charge density changes. Additionally, the localized nature of the electron density changes suggests that coupling terms for $J(r > 2a_0)$ are unlikely.

This behavior contrasts with that observed in many other intercalation compounds, such as Li$_x$C\cite{persson2010thermodynamic,safran1980electrostatic} and Li$_x$V$_2$O$_5$\cite{braithwaite1999lithium}, where the changes in charge density are not localized about intercalants and long-range electrostatic interactions are only partially screened. The behavior observed here is more similar to that seen in Li$_{x}$CoO$_2$ \cite{ceder1997application, ceder1999phase}, wherein a screening of electrostatic interactions between Li cations was attributed to nearby O orbitals. However, the charge density differences observed in our case (Fig. 3) are much more localized around the intercalants. This can be rationalized by the increased ability of Sc and Zn to hybridize with and polarize this host lattice owing to their higher oxidation states and larger orbitals. The host lattice itself is also relatively soft. Accordingly, the, the S \textit{p} orbitals surrounding the intercalants polarize to form a large shell of diffuse charge density. This is reflected in the DOS (Fig. 2), which suggests some hybridization of the S \textit{p} orbitals, as well as in the spatial distribution of charge density differences (Fig. 3). Similar behavior is observed in DFT charge density studies of other Zn-intercalated TMDs \cite{zhang2021metal, yang2023unlocking, wu2020ultrathin, tang2023unravelling}.

With this physical origin in mind, we present the coupling constants $J(r)$ obtained from fitting a series of DFT calculations for different intercalant configurations at varying values $x$ (see SI for details). The resulting $J(r)$ values, along with their associated root mean squared error, are given in Table 1. For Zn intercalation, the fit is reasonable, with an RMSE of 9 meV, allowing for approximate coupling constants and a corresponding phase diagram to be derived. However, for Sc-intercalated TaS$_2$, the RMSE is much larger (52 meV), making quantitative agreement with experimental data difficult. The significantly larger $J(a_0)$ for Sc, a higher oxidation state ion, is expected due to the stronger repulsive electrostatic interactions persisting at inter-site distances of $a_0$ in the Sc system. 

We believe the main source of uncertainty in the model stems from the assumption of pairwise coupling, which becomes less accurate for higher oxidation state ions. Since the coupling is not purely electrostatic, a pairwise model breaks down as additional charge transfer occurs and the host charge distribution becomes more diffuse. Consequently, interactions beyond pairwise coupling likely play a more prominent role in Sc-intercalated structures due to increased charge transfer, greater hybridization, and the marginally smaller lattice constant (1.8\% smaller than for Zn). 

Furthermore, we note that deviations from the lattice model are expected when the extent of charge transfer per ion varies with intercalation density. As we see more variability in the Sc Bader charges computed at different $x$, the presumed $x$-independence of $\mu$ may also contribute to a poorer agreement between the lattice model and DFT in the Sc-intercalated structure. While implementing a cluster model is beyond the scope of this paper, we will proceed with the obtained fits for both structures to capture the general trends in the superlattice phase diagram for both systems. In addition, the relatively good fit obtained for Zn-intercalated TaS$_2$ suggests that the pair-wise model is an increasingly well-motivated assumption for smaller oxidation state ions, which may prove helpful for further work characterizing a more diverse range of intercalation compounds.

\begin{table}[h]
\caption{Fitting constants for rigid \textit{2H}-M$_x$TaS$_2$ (M = Zn, Sc) within the optimal geometry for full intercalation ($x=0.5, f=1$) of the bilayer interface. Additional structural and fitting details are in the SI. All values are in eV per M$_{2x}$Ta$_2$S$_4$ unit. 22 super-cell configurations were used across a range of intercalation densities. The larger RMS obtained for the Sc-intercalation is expected, and the inability to accurately model this system with a simple pairwise lattice model is rationalized in the main text. We attribute the unexpected difference in obtained $E_0$ between the two materials to this appreciable RMS.  
}
\begin{tabular}{ c c c c c c c }
\textrm{M}&
\textrm{$E_0$ }&
\textrm{$\mu$ }&
\textrm{$J(a_0)$ }&
\textrm{$J(\sqrt{3}a_0)$ }&
\textrm{$J(2a_0)$ }&
\textrm{RMS }\\
\hline
 Zn  & -49.131 & -2.756 & 0.206 & 0.019 & 0.006 & 0.009  \\ 
\hline
Sc  & -49.206 & -9.672 & 0.281 & 0.019 & 0.006 & 0.052  \\ 
\end{tabular}
\end{table}

\begin{figure*}
\includegraphics[width=16cm]{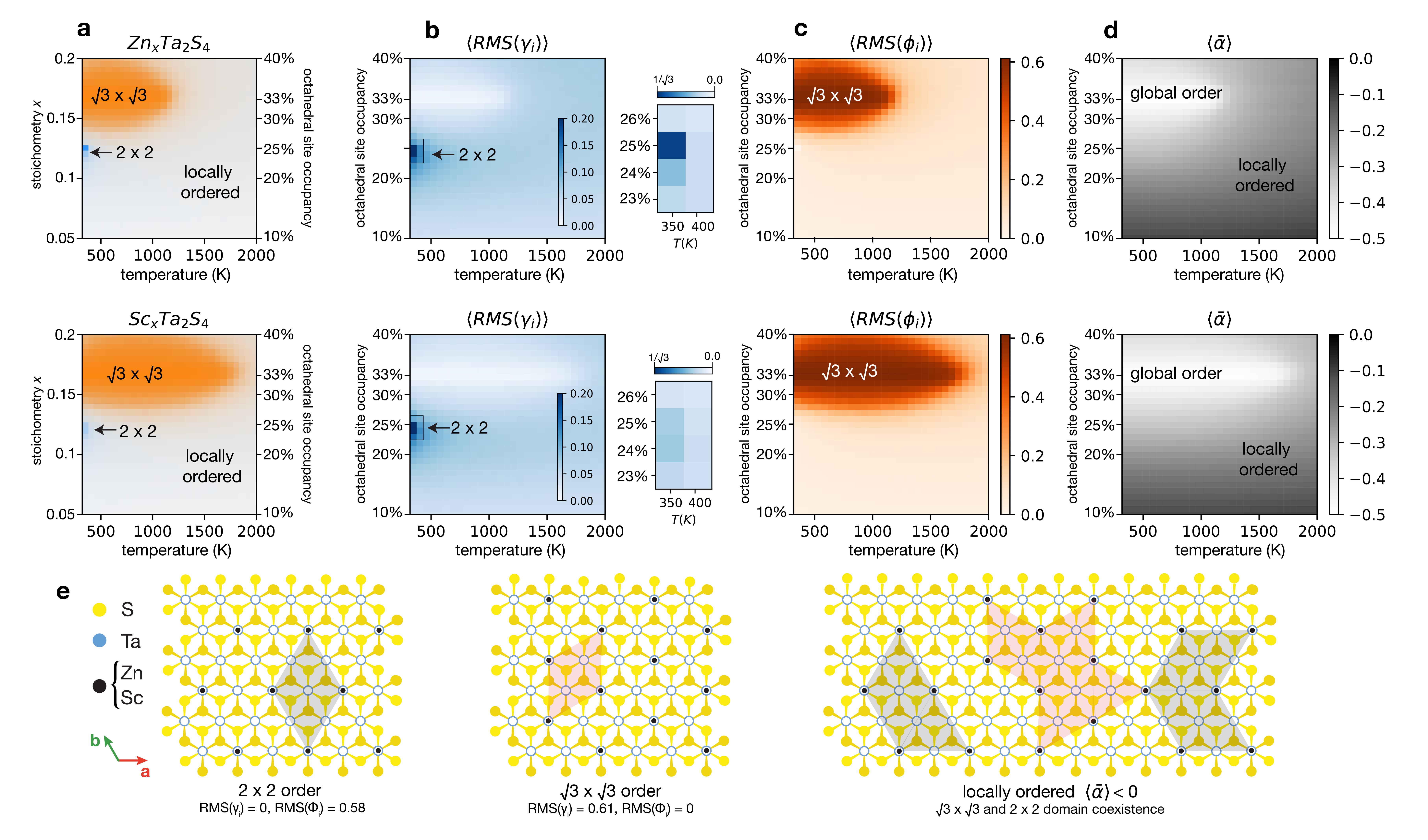}
\caption{ (a) Temperature and intercalation density phase diagram for a next-next-nearest neighbor triangular lattice gas model with coupling constants representative of a Zn-intercalated \textit{2H}-TaS$_2$ bilayer (top) and a Sc-intercalated \textit{2H}-TaS$_2$ bilayer (bottom). (b-c) Corresponding thermally averaged $RMS(\phi_i)$ and $RMS(\gamma_i)$ from a 324 lattice site simulation with periodic boundary conditions used to identify long-range order. Traces illustrating convergence are within the SI. (d) The average value for the local order parameter, with negative values indicating short-range order with coexisting \two and \three phases and avoidance of nearest neighbor occupancy. Further details and traces illustrating convergence are provided in the SI. (e) Occupation schematics characteristic of each ordered phase.}
\end{figure*}

\begin{figure}
\includegraphics[width=12cm]{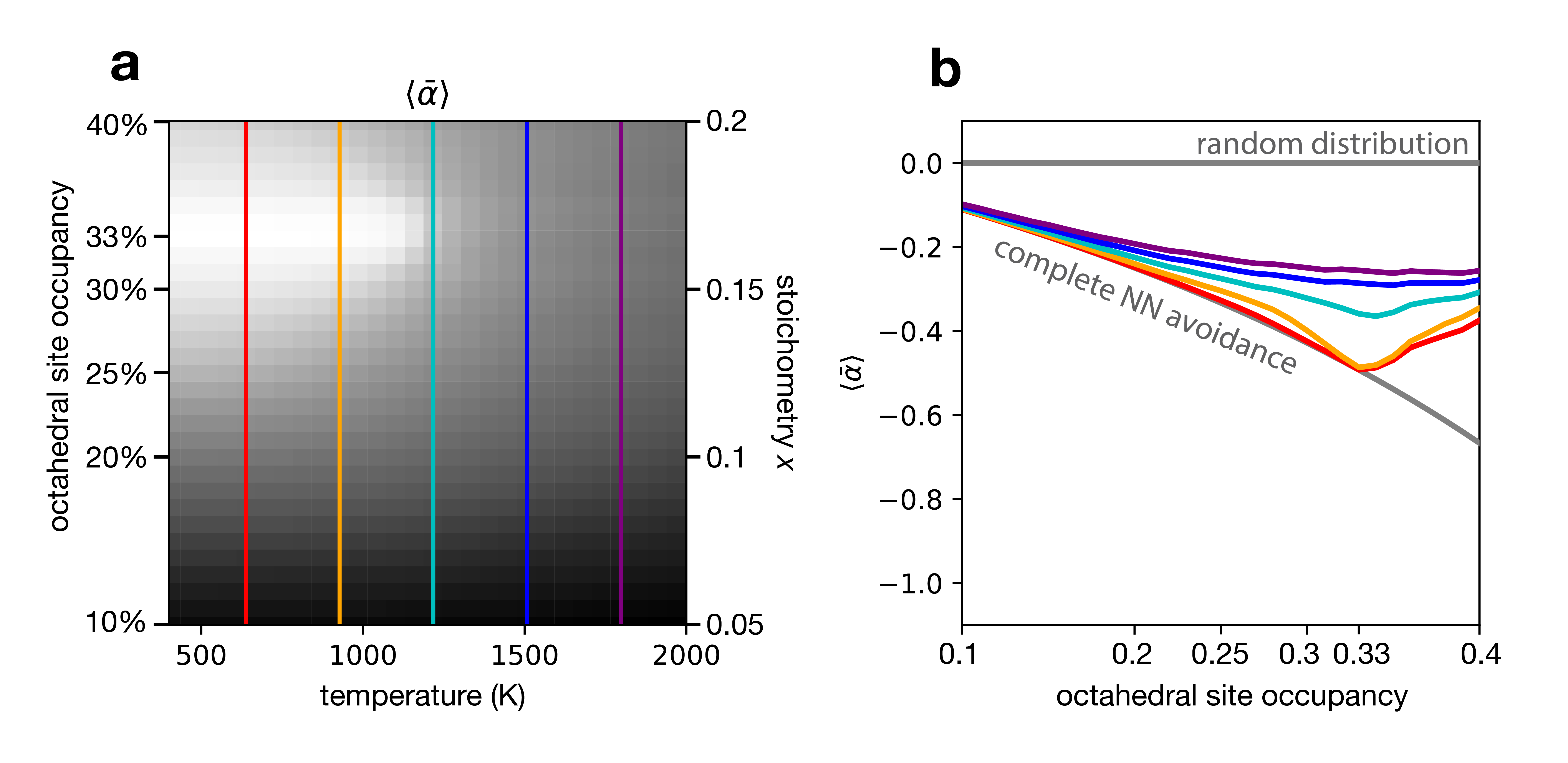}
\caption{ Line-cuts displaying $\langle\bar{\alpha}\rangle$ vs $x$ for a series of temperatures for the Zn-intercalated \textit{2H}-TaS$_2$ bilayer, compared to the $\langle\bar{\alpha}\rangle = 0$ expected for randomly distributed intercalants and the minimal $\langle\bar{\alpha}\rangle = -f/(1-f)$ value associated with complete nearest neighbor ($r=a_0$) avoidance, obtainable only for $f$ $\leq 0.33$ where $f$ is the fraction of occupied octahedral sites, equal to $x/2$ for an intercalated bilayer. 
} 
\end{figure}

\subsection{Superlattice Phase Diagram}

We next investigate the thermodynamically optimal phase diagram associated with these lattice gas parameters using Monte Carlo simulations. The resulting temperature and intercalation density phase diagram is shown in Fig. 5, presented in terms of the previously described order parameters, $RMS(\phi_i)$, $RMS(\gamma_i)$, and $\langle\bar{\alpha}\rangle$. Plots of each order parameter are shown separately to identify the long-range \two order (Fig. 5b), long-range \three order (Fig. 5c), and short-range avoidance of nearest neighbor occupation (Fig. 5c). Since $J(a_0)$ is at least an order of magnitude larger than $J(r > a_0)$, we anticipate local ordering to persist when $k_B$T is less than the $J(a_0)$ term, corresponding to temperatures of roughly 2400 K and 3300 K for Zn and Sc respectively. Consequently, all relevant crystal growth temperatures (below approximately $\sim$1000--1400 K) \cite{hulliger_magnetic_1970, van_laar_magnetic_1971, Parkin1980a, husremović2022hard} should exhibit local ordering. Accordingly, line-cuts of the short-range order parameter $\langle\bar{\alpha}\rangle$ (Fig. 6) illustrate that even in the disordered phase, $\langle\bar{\alpha}\rangle$ remains significantly below the zero value expected for random ion occupancy. 

At one-third interstitial site occupation ($x=1/6$ in this bilayer model, equivalent to $x=f=1/3$ for 3D crystals), the large energy penalty for nearest neighbor occupation favors long range \three order (Fig. 5e), as it is the only occupation that completely avoids $r=a_0$ interactions. This phase remains stable up to roughly 1200 K and 1700 K for Zn and Sc respectively as fluctuations incur substantial energy penalties associated with these $r=a_0$ interactions. Conversely, at one-quarter interstitial site occupation ($x=1/8$ in this bilayer model, equivalent to $x=f=1/4$ for 3D crystals), there is a relatively minor energy cost to disrupt \two order through forming locally-ordered \three, \two, and vacant domains which still avoid $r=a_0$ interactions (Fig. 5e).  As a result, \two order is only energetically favored over this higher-entropy phase by a slim margin, persisting over a smaller range of intercalation densities and at relatively low temperatures.

These results suggest that experimentally targeting bulk T$_{1/4}M$S$_2$ materials with exclusively \two structures, which are of great interest for non-relativistic spin splitting in compensated magnetic systems \cite{mandujano2024itinerant, regmi2024altermagnetism, lawrence2023fe} (so-called altermagnetism \cite{vsmejkal2022beyond}), likely requires both near-perfect stoichiometry and slow cooling down to the ambient temperatures where long-range \two is stable and kinetically trapped. In contrast, long-range \three order persists over a broader range of $x$ and is stable up to roughly 1200 K and 1700 K for Zn and Sc respectively. This phase generally requires less stringent cooling conditions, though cooling rate remains a factor in influencing long-range disorder, including heterochirality, in these materials.\cite{goodge2023consequences}

\section{Conclusion}
We present an extended pairwise lattice model for Zn and Sc ordering in bilayer \textit{2H}-TaS$_2$, derived from DFT calculations, and investigate its physical interpretation. Our results highlight that electrostatic interactions are largely screened to yield inter-site interactions that predominantly take place between nearest and next-nearest neighboring sites. This model is used to obtain a temperature and intercalation density phase diagram for superlattice ordering, offering critical insights into the approximate phase diagram of divalent and trivalent ion ordering at the vdW interface of \textit{2H}-TaS$_2$ and other related TMDs like NbS$_2$. The resulting phase diagram reveals a strong preference for \three long-range order and coexisting locally ordered \three and \two domains, even at intercalation densities corresponding to a quarter of the octahedral interstitial sites being filled. 

We anticipate that this approach to modeling lattice occupancy will serve as a useful reference for targeting key superlattice phases experimentally, and provide a foundation for formulating the more sophisticated models needed to capture the rich range of magnetic and electronic properties observed in transition metal intercalated materials. The role of out-of-plane elastic effects, which were not included here, may be substantial and should be considered in future work. Future studies may also explore how out-of-plane intercalant order/disorder in 3D crystals contributes to the overall superlattice ordering of these materials. 

\begin{acknowledgement}
We acknowledge helpful conversations with Lilia Xie, Oscar Gonzalez, Katherine Inzani, and Mark Asta. This work was funded by the U.S. Department of Energy, Office of Science, Office of Basic Energy Sciences, Materials Sciences and Engineering Division under Contract No. DE-AC02-05-CH11231 within the Theory of Materials program. Computational resources were provided by the National Energy Research Scientific Computing Center and the Molecular Foundry, DOE Office of Science User Facilities supported by the Office of Science, U.S. Department of Energy under Contract No. DEAC02-05CH11231. The work performed at the Molecular Foundry was supported by the Office of Science, Office of Basic Energy Sciences, of the U.S. Department of Energy under the same contract.
 I.M.C. acknowledges a pre-doctoral fellowship award under contract FA9550-21-F-0003 through the National Defense Science and Engineering Graduate (NDSEG) Fellowship Program, sponsored by the Office of Naval Research (ONR).
\end{acknowledgement}

\subsection*{Data and Code Availability}
The Monte Carlo implementation and VASP files are available at within the SimpleLatticeMC and SuperlatticePhaseSpace\_2H-TaS2 directories respectively within github.com/Griffin-Group. 

\subsection*{Author Contributions}
I.M.C., D.T.L., D.K.B. and S.M.G. conceived the study. I.M.C and B.J.K performed DFT. I.M.C implemented and performed MC, and analyzed and interpreted all data. S.M.G and D.K.B supervised the work. I.M.C and S.M.G wrote the manuscript with input from all co-authors.

\subsection{Supplementary Section 1: Computational Details and Functional Comparison}

 We performed Density Functional Theory (DFT) calculations within the Vienna Ab Initio Simulation Package (VASP) \cite{hafner2008ab}. We used projector augmented wave (PAW) pseudopotentials \cite{blochl1994projector} containing valences of $3d^{10}4s^2$, $3s^23p^63d^14s^2$, $5p^65d^36s^2$ and $3s^23p^4$ for Zn, Sc, Ta, and S respectively. All DFT calculations were performed at the GGA+U level using the PBEsol \cite{perdew2008restoring} functional and a Dudarev-type on-site Hubbard correction \cite{dudarev1998electron} on the Zn and Sc d-orbitals of U=3.0 eV and U=2.9 eV (unless otherwise noted), obtained via linear response \cite{kulik2006density, cococcioni2005linear} using the bulk structural configuration. The U value was obtained by optimizing the geometry and obtaining the optimal U a few times until convergence with an energy cutoff of 600 eV. PBEsol was selected due to its ability to accurately predict the experimental structure of bulk \textit{2H}-TaS$_2$ when compared to other commonly used GGA candidates, shown in Supplementary Table 1. All systems were metallic for all $x$ choices. 

Calculations of Zn$_x$TaS$_2$ and Sc$_x$TaS$_2$ were both performed using an energy cutoff of 800 eV and a vacuum height of 21 \textrm{\AA} for slabs, with a vanishing local potential between slabs. We used a 23/23/2 and 23/23/6 $\Gamma$-centered k-point grid for bilayer and bulk Ta$_{2}$S$_{4}$, scaled for each super-cell for the occupancy fit and DOS, which corresponds to a length cutoff of 33 \textrm{\AA}. $\Gamma$-centered k-point grids of 13/13/2 and 13/13/4 (19 \textrm{\AA} cutoff) were used for bilayer and bulk structural optimization and U determination. When values are compared for various U values, each calculation was re-initialized randomly to avoid possibly spurious charge density similarity, as gradually shifting U can target local minima. \cite{smolyanyuk2024fragility}

Structural optimizations were converged to a force tolerance of 1 meV/\textrm{\AA} throughout. Geometry optimization and U determination calculations were performed using first-order Methfessel-Paxton (with $\sigma = 0.2$ eV corresponding to an entropy of $\approx 0.1$ meV per atom). Electronic densities of states (DOS) were calculated with tetrahedron smearing, with a Gaussian smoothing of $\sigma=0.03$ included for visualization. Orbital projections were performed by projecting onto the PAW sphere, with the x and y axis defined parallel to a and c respectively. Projections onto spherical harmonics using Bader radii are additionally given in Supplemental Fig. 3.
 
Spin-orbit coupling and spin polarization were excluded throughout. The inclusion of spin polarization (starting from an initialization of one Bohr magneton per atom) relaxed to a non-magnetic ground state for both bilayer Sc$_x$Ta$_2$S$_4$ or Zn$_x$Ta$_2$S$_4$. Bader charge calculations were obtained using the referenced algorithm \cite{tang2009grid}, using the core charge density as a reference and fast Fourier transform grids of $192 \times 192 \times 1920$ for the M$_{x}$Ta$_{2}$S$_{4}$ and $480 \times 480 \times 4800$ for the M$_{9x}$Ta$_{18}$S$_{36}$, sufficient to capture the expected total charge. 

Equilibrium Monte Carlo (MC) simulations were performed using a user-written Julia implementation of Kawasaki dynamics, available at github.com/Griffin-Group/SimpleLatticeMC. We note in the implementation that for detailed balance to be satisfied for Kawasaki dynamics, we must select random hop directions indiscriminately and reject the move when such a move is incompatible with the current lattice occupation. MC simulations used periodic boundary conditions within an 18x18 simulation cell using 2x10$^4$ sweeps after an equilibration period of 5x10$^3$ sweeps, sufficient for convergence in these relatively high-temperature simulations as seen in the order parameters traces (Fig. 5). An equilibration period of 5x10$^4$ and collection period of 5x10$^5$ sweeps were used for more slowly convergent regions ($ T < 410 K$ for $f=0.24-0.26$ of both Zn and Sc) and $4.95 \times 10^6$ sweeps were used for ($ T < 350 K$ for $f=0.25$ for Zn), as seen traces in Figs. 6-8. A rigorous evaluation of finite size effects with, for example, a Binder coefficient analysis, is not included in this work, primarily because the coupling constants are not known to a high enough degree of uncertainty. 

\begin{figure}
\includegraphics[width=8cm]{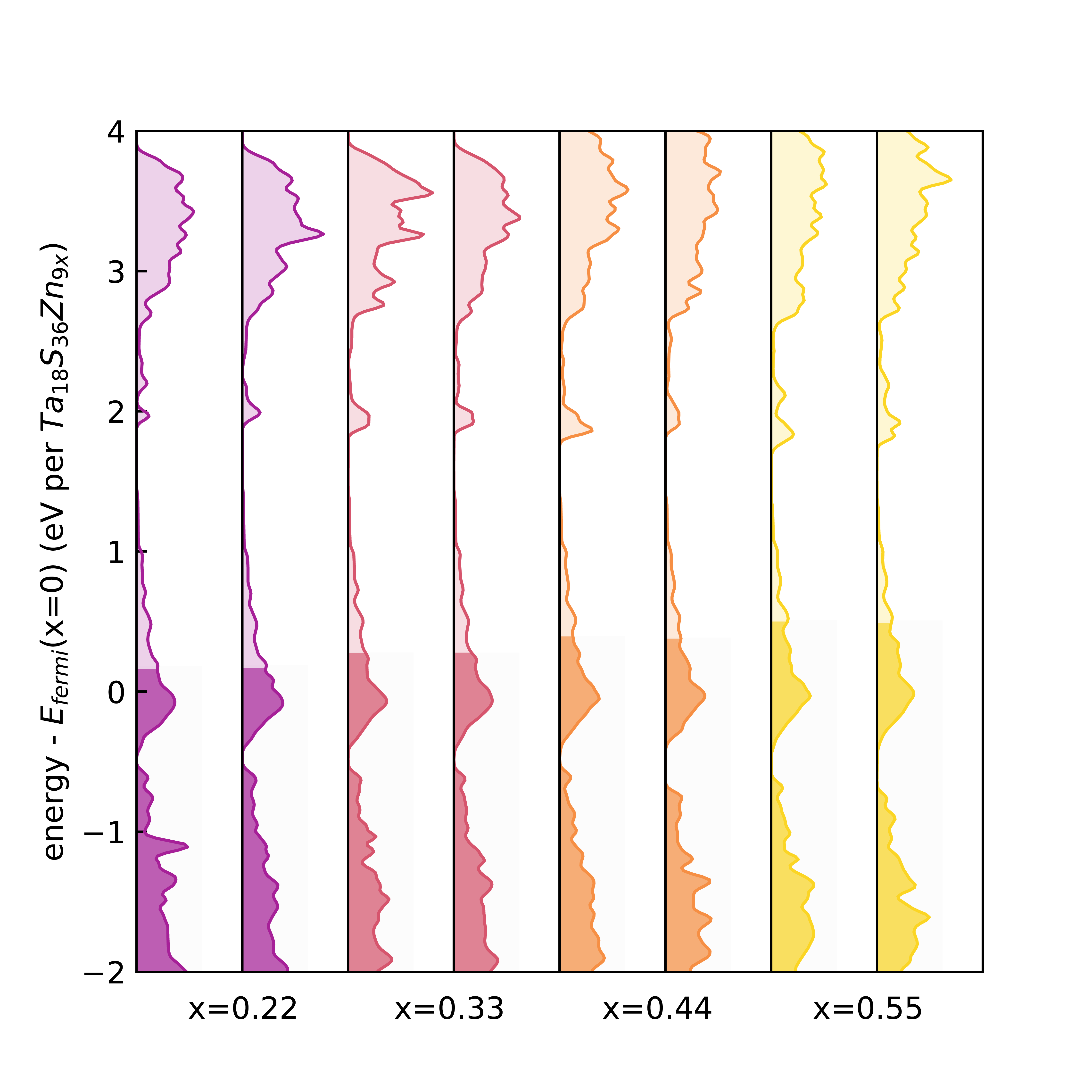}
\caption{ \textbf{Figure Appendix 1:} Calculated electronic density of states (DOS) projected onto the PAW sphere for Ta$_{18}$S$_{36}$Zn$_{9x}$ at a series of intercalation densities, $x$, and various intercalant arrangements. Fermi energies are shown as fill lines and all energies are shown relative to the Fermi energy of the bare host, $x=0$. A Gaussian filter with width $\sigma =0.03$ is applied before visualizing.} 
\end{figure}

\begin{figure}
\includegraphics[width=8cm]{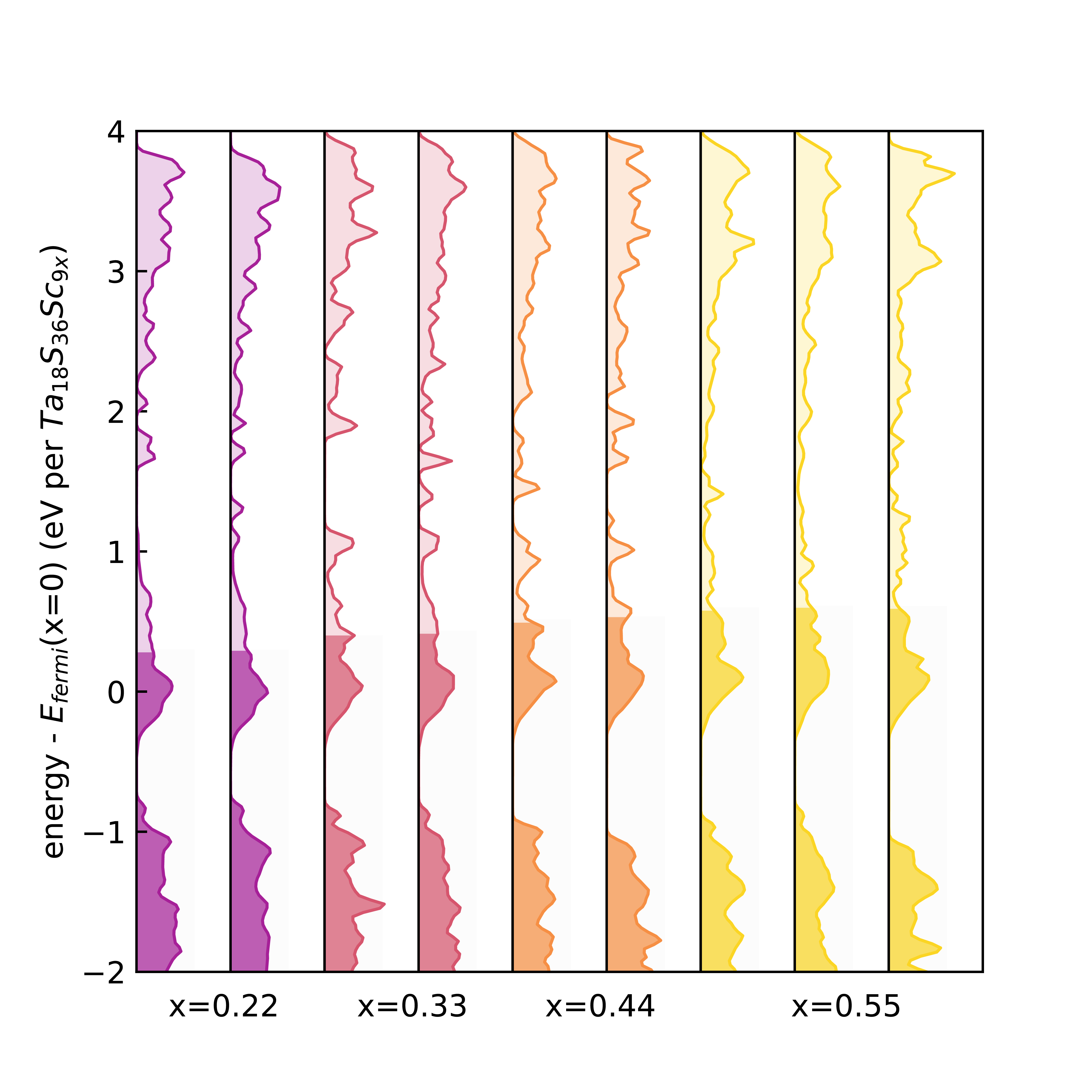}
\caption{ \textbf{Figure Appendix 2:} Density of states projected onto the PAW sphere for Ta$_{18}$S$_{36}$Sc$_{9x}$ at a series of intercalation densities, $x$, and various intercalant arrangements. Fermi energies are shown as fill lines and all energies are shown relative to the Fermi energy of the bare host, $x=0$. A Gaussian filter with width $\sigma =0.03$ is applied before visualizing} 
\end{figure}

\begin{figure}
\includegraphics[width=8cm]{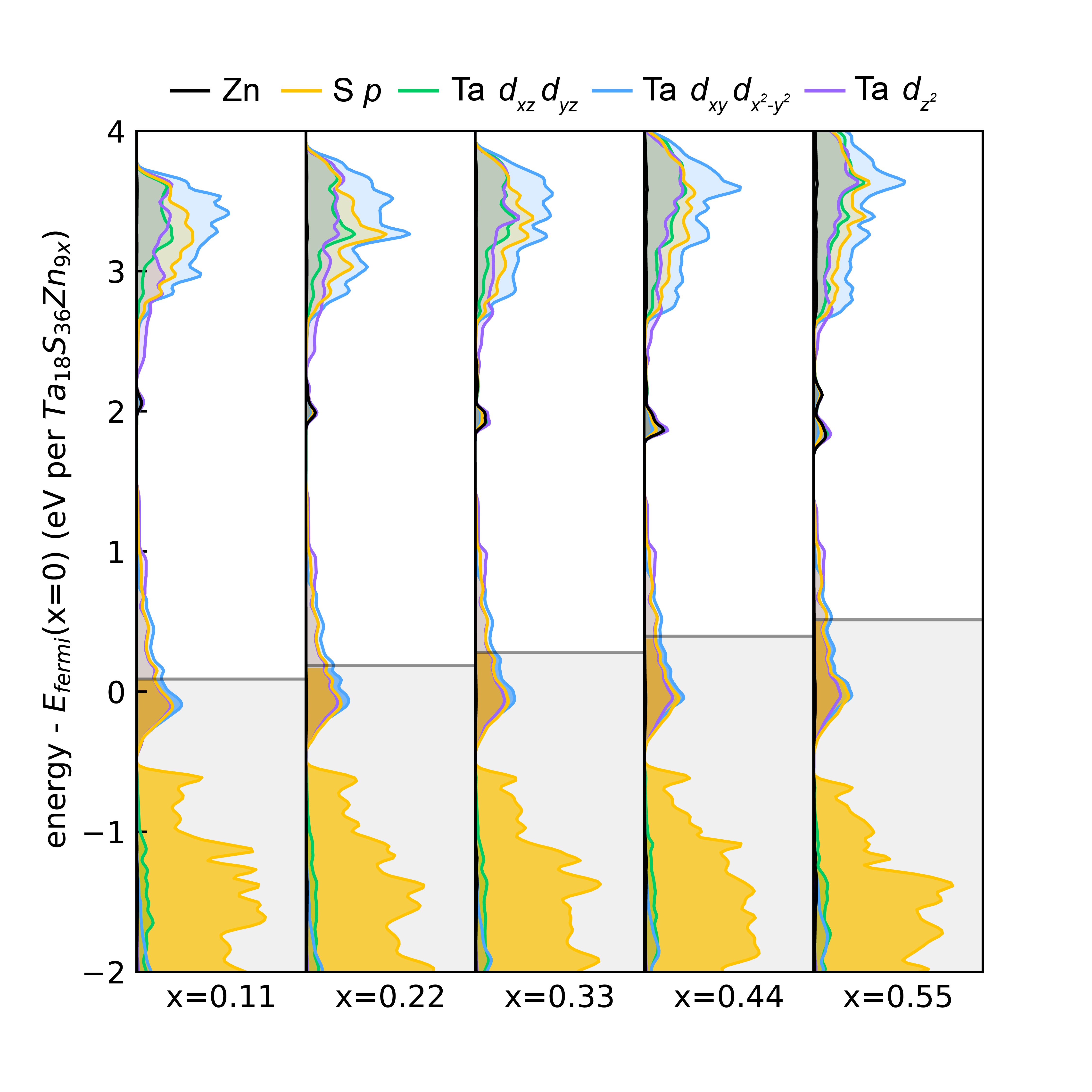}
\caption{ \textbf{Figure Appendix 3:} Density of states projected onto Bader charge sphere for Ta$_{18}$S$_{36}$Zn$_{9x}$. Fermi energies are shown as grey lines and all energies are shown relative to the Fermi energy of the bare host, $x=0$. A Gaussian filter with width $\sigma =0.03$ is applied before visualizing.} 
\end{figure}

\begin{table}[h]
\caption{Relaxed structures of \textit{2H}-TaS$_2$ (bulk $P6_3/mmc$). * denote bilayer slab structures, while those unlabeled are bulk. An energy cutoff of 500 eV was used. }
\begin{tabular}{ l l l l l }
\textrm{}& \textrm{Functional}& \textrm{a (\AA) }& \textrm{c (\AA)}& \textrm{VdW (\AA)} \\
 \hline
 \textit{2H}-TaS$_2$               & Exp         & 3.314 & 12.097 & 2.918 \\
 \textit{2H}-TaS$_2$               & DFT-D3      & 3.320   & 12.073 &  2.909 \\
 \textit{2H}-TaS$_2$               & DFT-TS      & 3.313   & 11.673 &  2.713 \\
 \textit{2H}-TaS$_2$               & PBE         & 3.340   & 13.963 &  3.856 \\
 \textit{2H}-TaS$_2$               & PBEsol      & 3.294   & 12.106 &  2.932 \\
 \textit{2H}-TaS$_2$               & optB86b-vdw & 3.317   & 12.061 &  2.891 \\
 2H-TaS$_{2}^*$             & PBEsol      & 3.296   & 12.054 &  2.916 \\
 2H-TaS$_{2}^*$             & DFT-D3      & 3.325   & 12.090 &  2.928 \\
\end{tabular}
\end{table}

\subsection{Supplementary Section 2: Fit to Lattice Hamiltonian}

Obtaining the lattice Hamiltonian amounts to finding the least squares solution to an over-determined set of linear equations for the inter-site interactions. This was accomplished using 1x1, 2x2, and 3x3 super-cells of variable site occupations, resulting in the following coupling between the ions on each site for the 1x1 and 2x2 super-cells shown below, defined analogously for the 3x3.
\begin{align*}
H_{ij}^{11} = 
\begin{bmatrix}
J_{00}^{11} \\
\end{bmatrix}
H_{ij}^{22} = 
\begin{bmatrix}
J_{00}^{22} & J_{10}^{22} & J_{01}^{22} & J_{11}^{22} \\
.           & J_{00}^{22} & J_{10}^{22} & J_{01}^{22} \\
.           & .           & J_{00}^{22} & J_{10}^{22} \\
.           & .           & .           & J_{00}^{22} \\
\end{bmatrix}
\end{align*}

Here the occupancy variables $\sigma_{nm}$ at each lattice site $r=na_1 + ma_2$ within the super-cell were ordered as: $(\sigma_{00}, \sigma_{10} ... \sigma_{01} ... \sigma_{11} ...)$, and $J_{nm}^{ij}$ denotes the aggregate interaction constant between the ions at lattice sites separated by a distance of $r=na_1 + ma_2$ within an $(i,j)$ super-cell. Therefore from energies of sufficiently many occupations of the super-lattice, we can obtain $J_{nm}^{ij}$ from inverting $E = \sigma^T H \sigma$ (in agreement with the double counting conventions shown prior). These aggregate interaction constants consist of the following. 
\begin{align*}
J_{ij}^{kl} &= \sum_{nm} J(|(kn+i)a_1 + (lm+j)a_2|) 
\end{align*}
For example $J_{10}^{41}$ arises from the combinations below since the coupling between these two sites involves a few interactions between different periodic repetitions, with offsets of $(n,m)=-1,0$ corresponding to a $J(3a_0)$ interaction and so on such that the following holds. 
\begin{align*}
J_{10}^{41} &= \sum_{nm} J(|(4n+1)a_1 + ma_2|) \\
&= 2J(a_0) + 2J(\sqrt{3}a_0) + 4J(\sqrt{7}a_0) +2J(3a_0) + ...
\end{align*}
This yields the following when using a truncation of $r \leq 2a_0$, where we know by symmetry that $J_{nm}^{10} = J_{mn}^{01} = J_{mn}^{11}$ and $J_{nm}^{20} = J_{mn}^{02} = J_{mn}^{22}$.
\begin{align*} 
\begin{bmatrix}
J_{00}^{11} \\
J_{00}^{22} \\
J_{10}^{22} \\
J_{00}^{33} \\
J_{10}^{33} \\
J_{20}^{33} \\
J_{21}^{33} \\
\end{bmatrix}
= \begin{bmatrix}
6 & 6 & 6 \\ 
6 & 0 & 0 \\ 
2 & 2 & 0 \\ 
0 & 0 & 0 \\ 
1 & 0 & 1 \\ 
1 & 0 & 1 \\ 
0 & 3 & 0 \\ 
\end{bmatrix} 
\begin{bmatrix}
J(a_0) \\
J(\sqrt{3}a_0) \\
J(2a_0) \\
\end{bmatrix}
\end{align*}
In practice, the fitting process therefore amounted to inverting the following over-determined linear relationship between the DFT energies $E^{DFT}_i$ and the associated number of host lattice unit cells ($N_{Ta_2S_4}$), intercalants ($N_{Int}$), and the numbers of each inter-site interaction ($N_{J1}, N_{Jr3}, N_{J2}$) as determined from the occupation variables following the logic outlined above. This was accomplished with a Moore-Penrose pseudo-inverse. 
\begin{align*} 
\begin{bmatrix}
E^{DFT}_1 \\
... \\
E^{DFT}_n \\
\end{bmatrix}
= \begin{bmatrix}
N_{Ta_2S_4} & N_{Int} & N_{J1} & N_{Jr3} & N_{J2}\\
... \\
N_{Ta_2S_4} & N_{Int} & N_{J1} & N_{Jr3} & N_{J2}\\
\end{bmatrix} 
\begin{bmatrix}
E_0 \\
\mu \\
J(a_0) \\
J(\sqrt{3}a_0) \\
J(2a_0) \\
\end{bmatrix}
\end{align*}

\subsection{Supplementary Section 3: Elasticity Effects}

\begin{table*}[h]
\caption{Optimal geometries for bulk 2H-TaS2 ($P6_3/mmc$) and corresponding bilayer slab structures. Bilayer MTa$_2$S$_4$ structures correspond to placing M intercalants in all pseudo-octahedral sites between the two TaS$_2$ layers and none above or beneath them. Bulk M$_2$Ta$_2$S$_4$ structures correspond to placing M intercalants in all pseudo-octahedral sites. Results are presented with two choices of pseudopotentials for the Sc atoms, denoted as $Sc$ and $Sc_{sv}$ with valences of 3 and 11 respectively. $Sc_{sv}$ was used elsewhere throughout this work. }
\begin{tabular}{ c c c c }
\textrm{System}&
\textrm{$a_0$ (\AA)}&
\textrm{$r_{SS}$ along $\hat{c}$ (\AA)}&
\textrm{$r_{TaTa}$ along $\hat{c}$ (\AA)}\\
\hline
Bilayer TaS$_2$               & 3.30 &  2.916  & 6.025 \\
Bulk TaS$_2$                  & 3.29 &  2.932  & 6.053 \\
\hline
Bilayer ZnTa$_2$S$_4$ ($Zn$, U=3.0)  & 3.36      &  2.997  & 6.118 \\
Bulk Zn$_2$TaS$_2$ ($Zn$, U=3.0)     & 3.27      &  3.802  & 7.009 \\
\hline
Bilayer ScTa$_2$S$_4$ ($Sc$, U=3.0)           & 3.30 & 3.632 & 6.869  \\
Bulk Sc$_2$TaS$_2$ ($Sc$, U=3.0)              & 3.29 & 3.735 & 6.997  \\
\hline
Bilayer ScTa$_2$S$_4$ ($Sc_{sv}$, U=2.9)  & 3.30 & 3.590 & 6.838  \\
Bulk Sc$_2$TaS$_2$ ($Sc_{sv}$, U=2.9)     & 3.29 & 3.640 & 6.914  \\
\end{tabular}
\end{table*}

\begin{table}[h]
\caption{Force constants and in-plane lattice mismatches obtained from DFT structural optimizations. We define $2\Delta = L_{ov} - L_{vv}$ and $2\Delta' = L_{oo} - L_{ov}$ where $L_{oo}, L_{ov}, L_{vv}$ are the distances between two adjacent occupied intersitital pseudo-octahedral sites, one vacant and one occupied site, and two vacant sites respectively. Derivation of the elastic Hamiltonian assumes $\Delta \approx \Delta'$. Force constants $K$ are obtained from the completely intercalated MTa$_2$S$_4$ bilayer by obtaining energies via DFT structural optimizations at a series of fixed $L_{oo}$ (accomplished through pinning the intercalant coordinates and unit cell shape/volume and either obtaining the energy of this structure ($K$) or allowing the TaS$_2$ lattice to corrugate within this constraint ($K'$)). See Supplementary Fig. 4 for further details and the associated raw DFT and fit ($R^2 = 0.999$). We note that these values were only needed for Zn$_x$Ta$_2$S$_4$ due to the negligible in-plane lattice mismatch observed in TaS$_2$ following Sc intercalation and therefore negligible lattice-relaxation driven contribution to the occupancy phase space.  }
\begin{tabular}{ c c c c }
\textrm{System}&
\textrm{$\Delta+\Delta'$ (\AA)}&
\textrm{$K$ (eV\AA$^{-2}$)}&
\textrm{$K'$ (eV\AA$^{-2}$)}\\
\hline
Bilayer Zn$_x$Ta$_2$S$_4$ & 0.03       & 22.28 & 17.63 \\
Bilayer Sc$_x$Ta$_2$S$_4$ & $\approx 0$  & -     & -     \\ 
\end{tabular}
\end{table}

Further complicating the modeling is the role of deformations in the TMD host about the intercalants. While the structural deformation of TMDs about these large transition metal intercalants is unstudied to our knowledge, prior work \cite{frechette2019consequences} has investigated the relaxation dynamics of binary alloys within the context of generalized elasticity theory to account for lattice mismatch on the phase space of multiple component nano-crystals using an effective lattice model that can be straight-forwardly adapted to model local lattice deformations about intercalants. 

We use a coarse-grained lattice for a generic bilayer transition metal dichalcogenide containing octahedrally coordinated holes in which intercalants reside. In this model, deformations of the 2D triangular lattice bonds represent uni-axial strain in the underlying host lattice, which in elasticity theory, are represented by Young's modulus, which is isotropic in-plane due to the system's symmetry and therefore can be represented using a single spring constant $k$. This amounts to using the following effective Hamiltonian to describe the effects of structural deformation on the intercalant phase space.  
 \begin{align*}
E_{elastic}(\sigma) &= \frac{\epsilon}{2} \sum_{ij} V(|r_i-r_j|) \sigma_i \sigma_{j}  
\end{align*} 

In the above, $V(r)$ is an effective elastic coupling term obtained from generalized elasticity theory \cite{frechette2019consequences}. Its functional form and derivation are provided in the following section. $V(r)$ depends only on the lattice size and symmetry and $\epsilon=k\Delta^2/8$ depends on the system's spring constant $k$ and the difference in optimal (in-plane) lattice size about an intercalated and vacant interstitial site.

\begin{figure}
\includegraphics[width=8cm]{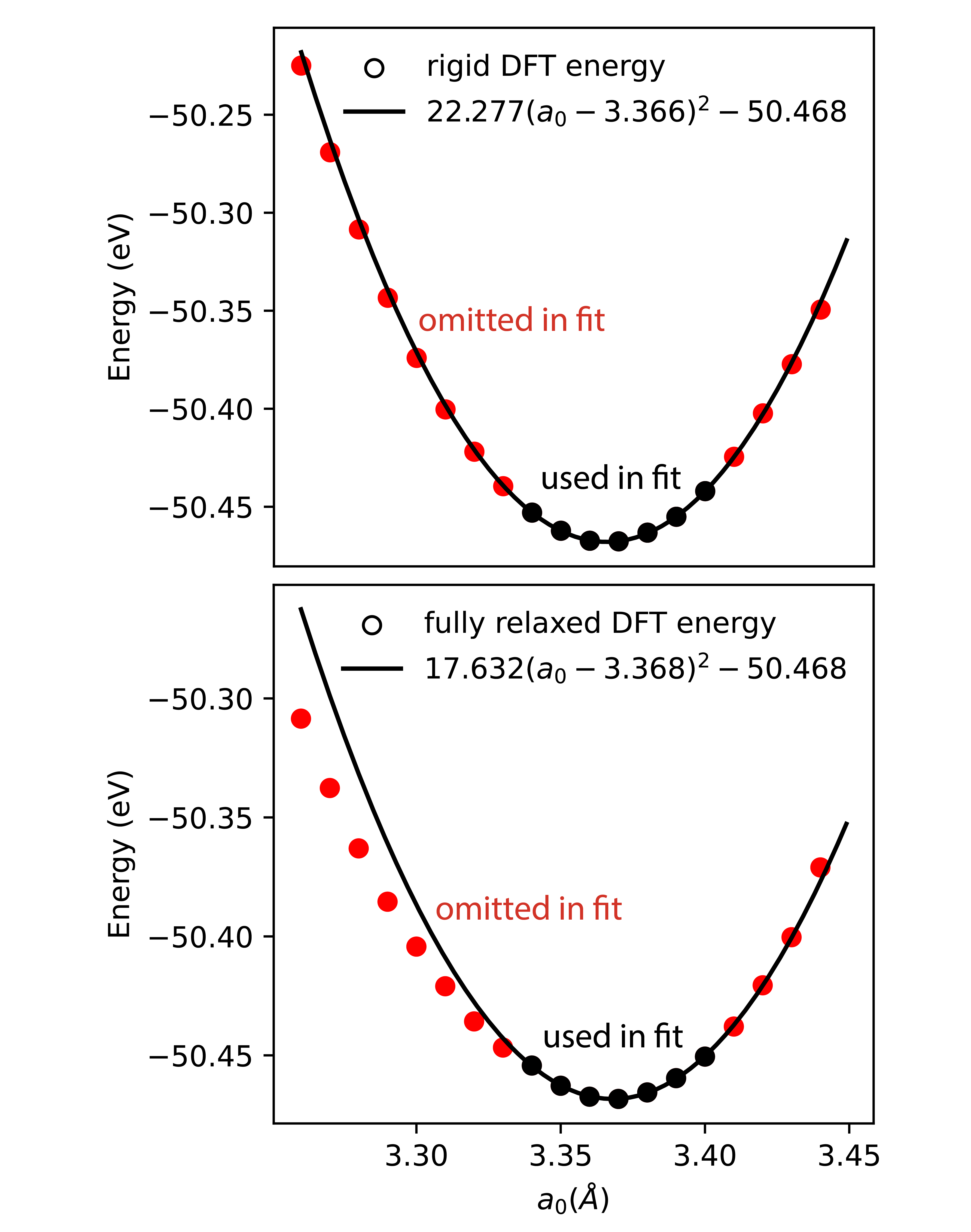}
\caption{ \textbf{Figure Appendix 4:} DFT energy as a function of $a_0$ for ZnTa$_2$S$_4$ and associated quadratic fit used in the determination of $k$. } 
\end{figure}

The spring constant $k$ (Fig. S4) and $\Delta$ values are given in Table 3. To match the conventions used in the following derivation, we multiply the spring constant by 4/6 (owing to the bond valency and fact that the sums are performed over all bond pairs i $\neq$ j rather than $i > j$), which yields 14.85 eV per square \AA. As there is only a very minor change in the in-plane lattice following intercalation (0.06\AA), so overall a relatively small elastic contribution to the phase space, with magnitude set by $\epsilon/2 = k\Delta^2/16 \approx 0.2$ meV. 

\begin{figure}
\includegraphics[width=16cm]{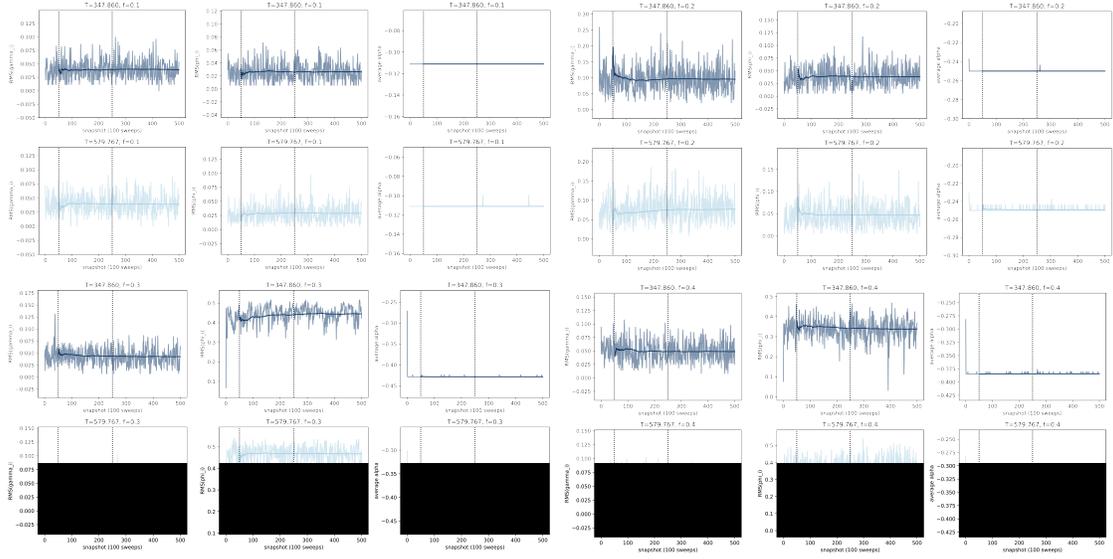}
\caption{ \textbf{Figure Appendix 5:} Monte Carlo traces and associated running averages for a select labeled series of temperature $T$ and site occupancy fraction $f$ to gauge convergence. The dotted lines correspond to the choice of warm-up sweeps ($5 \times 10^3$) and acquisition sweeps ($2 \times 10^4$) used in the phase space shown in the main text. Traces are shown for Zn intercalation only, where the same warm-up and collection periods were used for Sc intercalation at these points.} 
\end{figure}

\begin{figure}
\includegraphics[width=16cm]{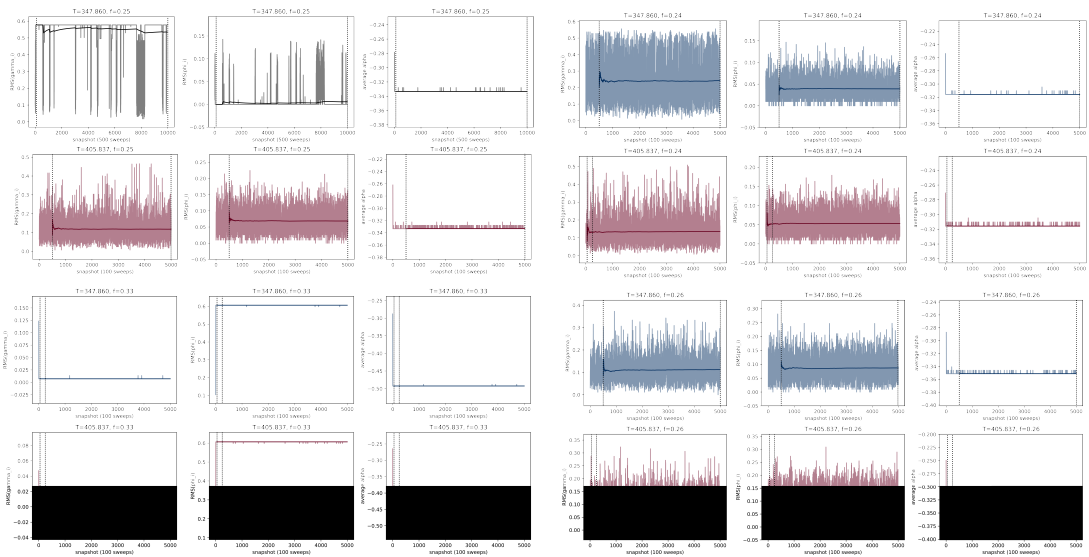}
\caption{ \textbf{Figure Appendix 6:} Monte Carlo traces and associated running averages for a select labeled series of temperature $T$ and site occupancy fraction $f$ to gauge convergence in more slowly convergent regions. The dotted lines correspond to the choice of warm-up sweeps ($5 \times 10^3$ or $5 \times 10^4$) and acquisition sweeps ($2 \times 10^4$, $5 \times 10^5$, or $4.95 \times 10^6$) used in the phase space shown in the main text.  Traces are shown for Zn intercalation. } 
\end{figure}

\begin{figure}
\includegraphics[width=16cm]{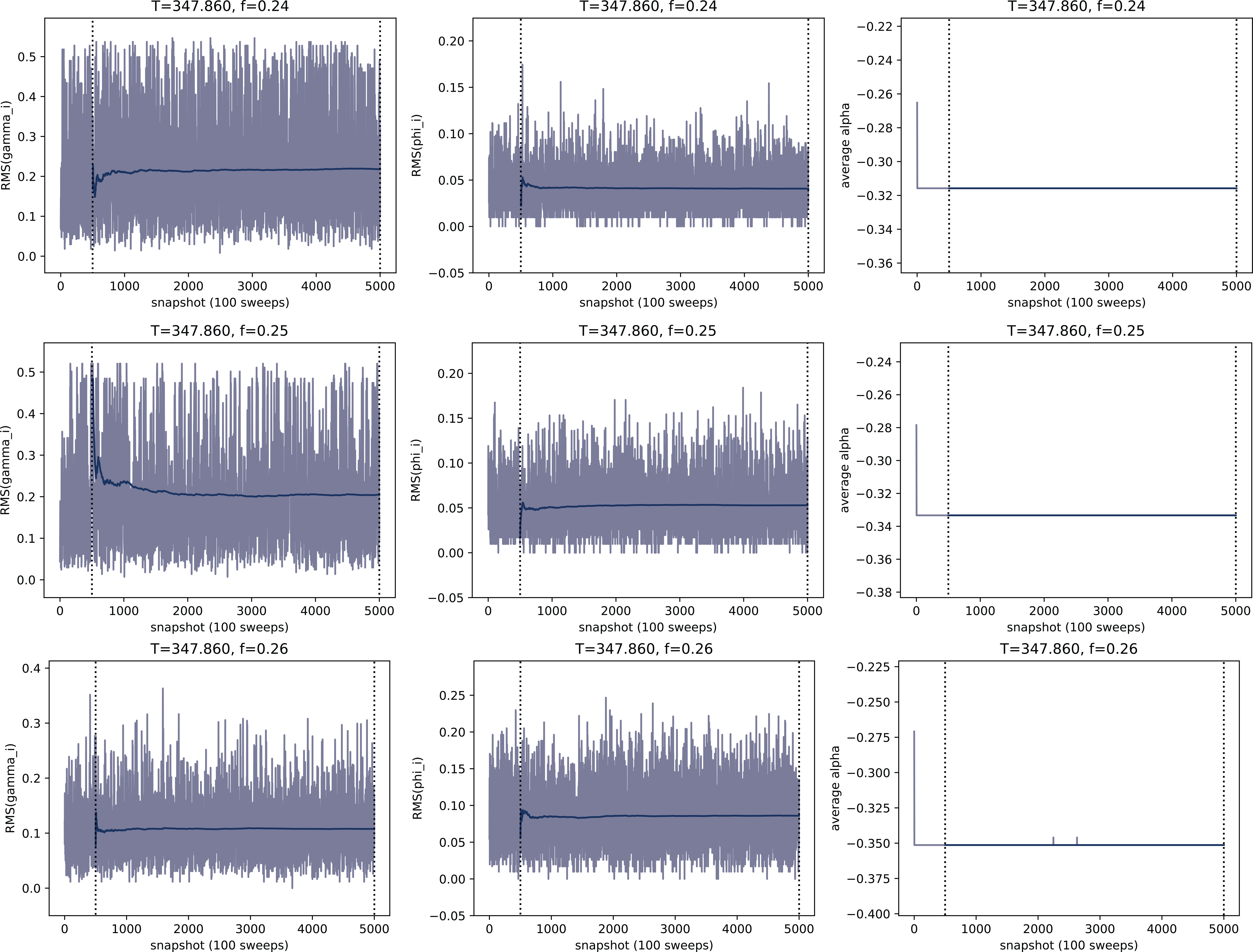}
\caption{ \textbf{Figure Appendix 7:} Monte Carlo traces and associated running averages for a select labeled series of temperature $T$ and site occupancy fraction $f$ to gauge convergence in more slowly convergent regions. The dotted lines correspond to the choice of warm-up sweeps ($5 \times 10^4$) and acquisition sweeps ($5 \times 10^5$) used in the phase space shown in the main text.  Traces are shown for Sc intercalation. } 
\end{figure}

\begin{figure}
\includegraphics[width=16cm]{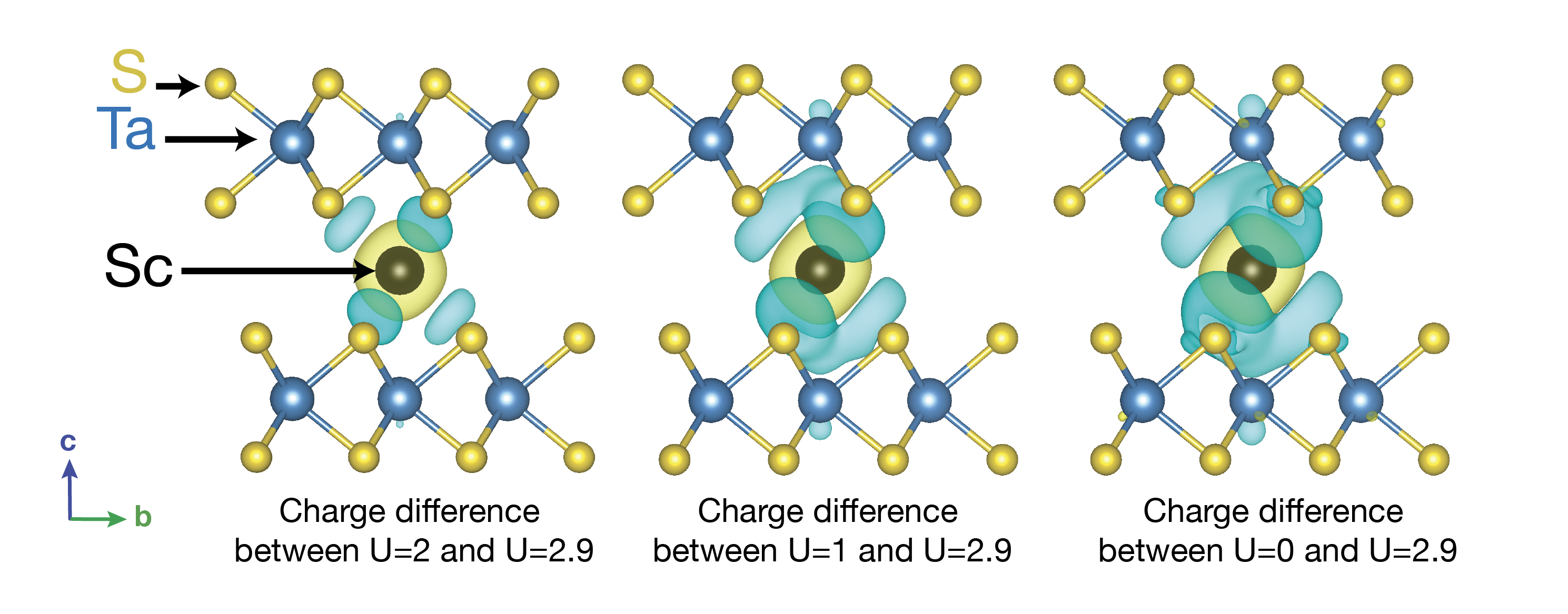}
\caption{ \textbf{Figure Appendix 8:} Impact of U value choice on the charge density of \textit{2H}-Sc$_{1/18}$TaS$_2$, which displayed a larger U value dependence in its Bader charge than Zn-intercalated \textit{2H}-TaS$_2$. Yellow and cyan isosurfaces reflect values of 0.0001 a.u. per Bohr radius cubed respectively. }
\end{figure}

\subsubsection{Supplementary Section 4: Elastic Energy Derivation}

We include here the derivation of the elastic Hamiltonian of \cite{frechette2019consequences}, designed for multi-component alloys. To derive this elastic Hamiltonian, they begin with the following model, in which the energy depends quadratically on the difference between bond lengths and their ideal values and a spring constant $k$ which sets the energy scale. They consider the case where coupling between occupation and displacement variables. enters in only through the variation in the ideal bond length $l(\sigma_R, \sigma_{R'})$ with occupation such that the general form of the elastic Hamiltonian as follows. 

\begin{align*}
H_e &= \frac{k}{4} \sum_{R, \alpha} (| a\alpha + v_R - v_{R+a\alpha} |  - l(\sigma_R, \sigma_{R+a\alpha}))^2
\end{align*}

In the above, the vector connecting nearest neighbor sites in the reference lattice is given by $a\alpha$, where $\alpha$ are the set of bond vectors for a hexagonal lattice and a is the in-plane lattice constant. The local displacements of the Rth site from its reference position are given by $v_R$, and the ideal distance between two lattice sites R and R' is given by $l(\sigma_R, \sigma_{R'})$. The binary variable $\sigma_R$ represents the occupation of a given lattice site in the spin basis such that $\sigma_R=-1$ for vacant sites and $\sigma_R=1$ for occupied sites. The dependence of $l(\sigma_R, \sigma_{R'})$ on the occupation variable accounts for the change in optimal lattice geometry following intercalation. They take the small lattice mismatch limit of the above Hamiltonian in order to integrate out the displacement field, for which \cite{frechette2019consequences} demonstrated good agreement with a direct application of Monte Carlo to the elastic equation above. They define $\Delta = (L_{ov} - L_{vv})/2$ and $\Delta' = (L_{oo} - L_{ov})/2$ in terms of the distances $L_{vv}$ between two vacant, $L_{oo}$ between two occupied, and $L_{ov}$ between a vacant and occupied site. This gives the following relationship for the ideal lattice length. 
\begin{align*}
l(\sigma_R, \sigma_{R'}) 
&= L_{vv} + \Delta (\sigma_{R}+\sigma_{R'}) + (\Delta' - \Delta) \sigma_{R} \sigma_{R'}
\end{align*}
For simplicity, one can assume that $\Delta=\Delta'$ so that $L_{ov} = (L_{oo} + L_{vv})/2$ to arrive at the following. 
\begin{align*}
H_{e} &= \frac{k}{4} \sum_{R, \alpha} ( a | \alpha +  \frac{1}{a} (v_R - v_{R+a\alpha}) | - L_{ov} - \frac{\Delta}{2}(\sigma_{R}+\sigma_{R+a\alpha}) )^2
\end{align*}
Using the linear (in $(v_R - v_{R+a\alpha})/a$) order approximation $ a | \alpha + \frac{1}{a} (v_R - v_{R+a\alpha}) | \approx 1 + \frac{1}{a} \alpha \cdot (v_R - v_{R+a\alpha}) + O((\frac{(v_R - v_{R+a\alpha})}{a})^2)$ and defining $s_R = \sigma_R - \frac{a - L_{ov}}{\Delta}$ and $u_R = \frac{v_R}{\Delta}$, the above simplifies to the following.
\begin{equation}
H_{e} = \frac{k\Delta^2}{4} \sum_{R, \alpha} (\alpha \cdot (u_R - u_{R+a\alpha}) - \frac{1}{2}(s_{R}+s_{R+a\alpha}) )^2
\end{equation}
This is quadratic in both the displacement and occupation variables with linear coupling between the two, allowing one to study lattice strain as linear response perturbation to the vacancy diffusion Hamiltonian. To integrate out the displacement field, one can exploit the translation symmetry of the coarse-grained lattice to exactly minimize equation 1 with respect to $\{ u_R \}$. Since the Hamiltonian is quadratic, minimizing with respect to $\{ u_R \}$ has the same effect as integrating it out apart from an additive constant. Due to the regular periodicity of the real space lattice, the binary variables can be represented by Fourier expansions within the lattice's first Brillouin Zone. (As $s_R$ is a discrete variable, this is formally not allowed, and doing so will introduce spurious high-frequency terms from the Gibbs phenomenon, however these will not affect the resulting formalism as potential variations of a wavelength smaller than the lattice spacing have no effect on this effective Hamiltonian).   
\begin{align*}
u_R = \frac{1}{N} \sum_q \hat{u}_q e^{-iq \cdot R} \text{\quad\quad} 
s_R = \frac{1}{N} \sum_q \hat{s}_q e^{-iq \cdot R} 
\end{align*}
In the above, $q = \sum_i n_i b_i / N_i $ where $n_i \in [0, N_i - 1]$ and $\{b_i\}$ are the reciprocal lattice vectors for the lattice. Since the Fourier basis functions are eigenstates of a translation operator, translationally invariant operators are diagonal in a plane wave basis. However since the Hamiltonian is only invariant with respect to a discrete set of translations by lattice vectors R, the plane waves need only be orthogonal when sampled at the real space lattice points (the plane waves can be aliased). This implies $\sum_R e^{-i(q+q')R} = N \delta_{q,q'}$ for the set of lattice vectors R. This yields the following Fourier space representation of equation 1, where $\hat{s}_q$ was replaced with $\hat{\sigma}_q$ since they are equivalent in all but the $q=0$ mode, for which the Hamiltonian is 0.
\begin{align*}
H_{e} = \frac{k\Delta^2}{4N} \sum_{q} \hat{u}_q \hat{F}_q \hat{u}_{-q} + \hat{\sigma}_{q} \frac{\hat{G}_q}{4} \hat{\sigma}_{-q} - \frac{i\hat{H}_q}{2} (\hat{u}_q \hat{\sigma}_{-q} - \hat{u}_{-q} \hat{\sigma}_q)
\end{align*}
The above expression uses the following matrices encoding details of the lattice structure, in which $\otimes$ denotes the Kronecker product.
\begin{align*}
\hat{F}_q &= \sum_{\alpha} \alpha \otimes \alpha (e^{-iqa\alpha} - 1) (e^{iqa\alpha} - 1) \\
\hat{G}_q &= \sum_{\alpha} \alpha (e^{-iqa\alpha} - 1) (e^{iqa\alpha} + 1) \\
i\hat{H}_q &= \sum_{\alpha} (e^{-iqa\alpha} + 1) (e^{iqa\alpha} + 1)
\end{align*}
Since the Hamiltonian is quadratic with respect to $\hat{u}_q$ and $\hat{F}_q$ and can be verified to be positive definite, the Hamiltonian has a single minimum with respect to $\hat{u}_q$. One can therefore minimize $H_{elastic}$ with respect to $\hat{u}_q$ to obtain an effective Hamiltonian depending only on the occupation variables, which has the same effect as integrating out the displacement field. As Fourier transforms of real functions have an even real part and odd imaginary part, $\hat{u}_{\pm q} = \hat{a}_q \pm i \hat{b}_q$ and $\hat{\sigma}_{\pm q} = \hat{c}_q \pm i \hat{d}_q$. This results in the following.
\begin{align*}
H_{e} = \frac{k\Delta^2}{4N} \sum_{q} \hat{F}_q (\hat{a}_q^2 + \hat{b}_q^2) + \frac{\hat{G}_q}{4} (\hat{c}_q^2 + \hat{d}_q^2) + \hat{H}_q (\hat{b}_q \hat{c}_q - \hat{a}_q \hat{d}_q )
\end{align*}
Minimizing $H_{elastic}$ with respect to $\hat{a}_q$ and $\hat{b}_q$ then leads to the following, which preserves the expected relationship for a real $u_R$. 
\begin{align*}
\hat{u}_{\pm q} = \frac{1}{2}(\hat{d}_q \mp i\hat{c}_q)\hat{F}_q^{-1}\hat{H}_q 
= \frac{\mp i}{2}\hat{\sigma}_{\pm q}\hat{F}_q^{-1}\hat{H}_q 
\end{align*}
The $\hat{u}_q$ defined above results in an effective elastic Hamiltonian of the form below where the effective potential $\hat{V}_q = \hat{G}_q - \hat{H}_q \hat{F}_q^{-1} \hat{H}_q$ and $\epsilon=k\Delta^2/8$
\begin{equation}
H_{effective} = \frac{\epsilon}{2N} \sum_{q} |\hat{\sigma}_q|^2 \hat{V}_q
= \frac{\epsilon}{2} \sum_{R,R'} \sigma_R V_{R-R'} \sigma_{R'}
\end{equation}
In a 2D triangular lattice, the bond vectors are given by 
\begin{align*}
\{\alpha_1, \alpha_2, \alpha_3, ... \} = \Bigg\{
\begin{pmatrix} \pm 1 \\ 0 \end{pmatrix},
\begin{pmatrix} \pm 1/2 \\ \pm \sqrt{3}/2  \end{pmatrix},
\begin{pmatrix} \pm 1/2 \\ \mp \sqrt{3}/2 \end{pmatrix}\Bigg\}
\end{align*}
This results in an effective potential of the following form, consistent with equation 4 of \cite{frechette2019consequences}. 
\begin{equation}
\hat{V}_q = \frac{4 (cos(aq_y) + 2 cos(\frac{\sqrt{3}a}{2}q_x)cos(\frac{a}{2}q_y) -3)^2 }{cos(\sqrt{3}aq_x) + (cos(aq_y)-2)(4cos(\frac{\sqrt{3}a}{2}q_x)cos(\frac{a}{2}q_y)-3)  }
\end{equation}

 \begin{figure*}
 \includegraphics[width=16cm]{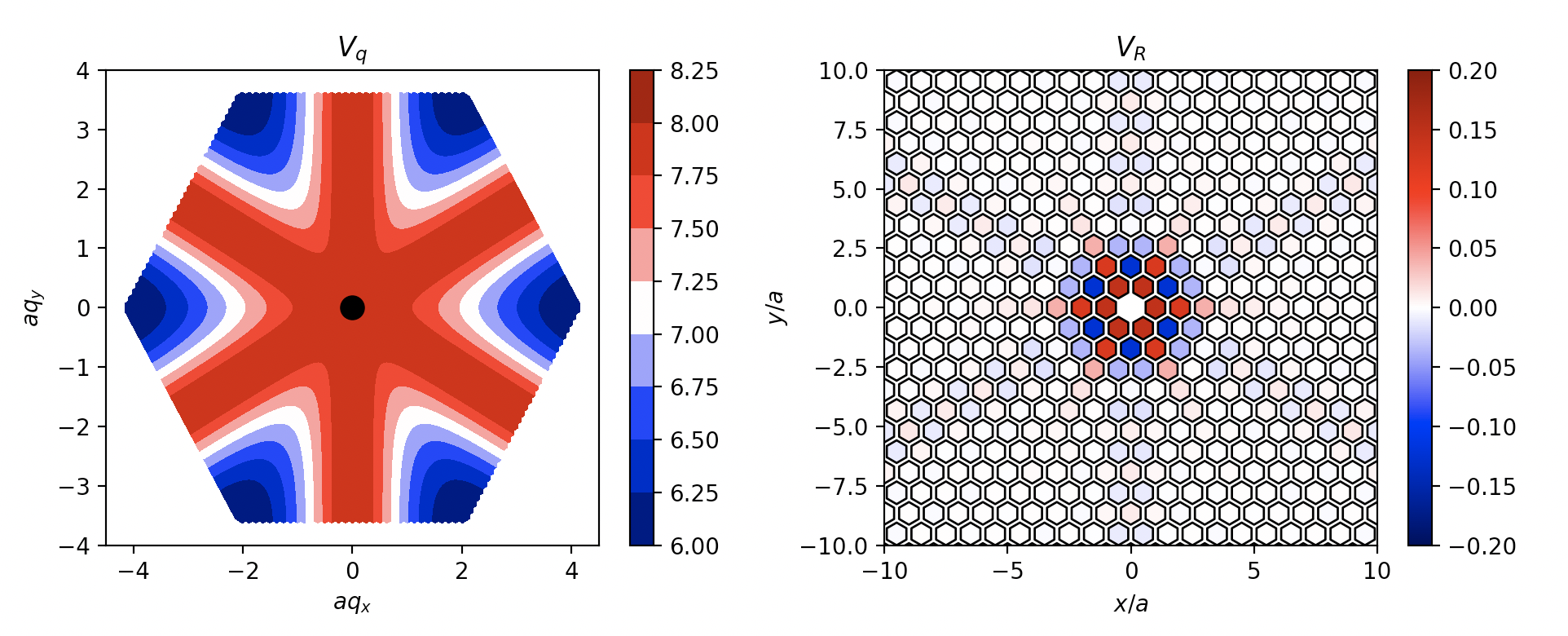}
 \caption{\textbf{Figure Appendix 9:} $V_q$ as given in equation 3 (left) and $V_R$ as given in equation 2 (right) evaluated as using an inverse Fourier transform within the first Brillouin zone with a momentum spacing of 0.005. The data is identical to that shown in Figs. 2a and 2b of the referenced work. \cite{frechette2019consequences} }
 \end{figure*}

$V_R$ can then be obtained by evaluating $V_R = \frac{1}{N} \sum_q \hat{V}_q e^{-iq \cdot R}$ within the first Brillouin zone and is shown in Supplementary Fig. 9. Using $\frac{\epsilon}{2} \approx 0.2$ meV from the prior section, the effective couplings between sites separated by distances of $a, \sqrt{3}a, 2a$ and $\sqrt{7}a$ are therefore approximately 3$\times 10^{-5}$, -2$\times 10^{-5}$, 2$\times 10^{-5}$, 4$\times 10^{-6}$ and eV respectively. While these elastic interactions decay much more slowly than the electrostatic interactions, they are at least 4 and 3 orders of magnitude less than the electrostatic interactions at $a$ and $\sqrt{3}a$ respectively. As discussed in the main text, the total impact of these long-ranged elastic effects also remains minor, as the associated phase diagram obtained from Monte Carlo for these interactions predicts ordering below $T \approx 4.6 K$ and therefore relatively minor impacts on the phase diagram presented (Fig. 5), with ordering transitions at temperatures 2-3 orders of magnitude higher. 

\bibliography{main}

\end{document}